\documentclass[aps,pra,reprint,floatfix,nofootinbib,superscriptaddress]{revtex4-2}

\usepackage{amsmath,amssymb,amsthm}
\usepackage{bm}
\usepackage{mathrsfs}
\usepackage{graphicx}
\graphicspath{{./}{source/}}
\usepackage{xcolor}
\usepackage{array,multirow,dcolumn,booktabs}
\usepackage{braket}
\usepackage{overpic}
\usepackage{tikz}
\usepackage{epstopdf}
\usetikzlibrary{arrows,arrows.meta,positioning,shapes,calc}
\usepackage[colorlinks=true,citecolor=blue,linkcolor=red,urlcolor=blue]{hyperref}

\newcommand{\PreserveBackslash}[1]{\let\temp=\\#1\let\\=\temp}
\newcolumntype{C}[1]{>{\PreserveBackslash\centering}p{#1}}
\newcolumntype{R}[1]{>{\PreserveBackslash\raggedleft}p{#1}}
\newcolumntype{L}[1]{>{\PreserveBackslash\raggedright}p{#1}}

\newcommand{\tcell}[2][3.2cm]{\parbox[t]{#1}{\raggedright #2}}
\begin{document}

\title{Geometric mode steering of the quantum Mpemba effect}
\author{Yingying Hong}
\affiliation{Department of Physics, Nanchang University, Nanchang 330031, China}
\author{Longxing Xu}
\affiliation{Department of Physics, Nanchang University, Nanchang 330031, China}
\author{Weiwei Zhang}
\affiliation{Department of Physics, Nanchang University, Nanchang 330031, China}
\author{Jie Ren}\email{xonics@tongji.edu.cn}
\affiliation{Center for Phononics and Thermal Energy Science, China-EU Joint Lab on Nanophononics, School of Physics Science and Engineering, Tongji University, Shanghai 200092, China}
\author{Jianhui Wang} \email{wangjianhui@ncu.edu.cn}
\affiliation{Department of Physics, Nanchang University, Nanchang 330031, China}

\begin{abstract}
The slowest Liouvillian mode often bottlenecks the relaxation of an
open quantum system to its steady state. Standard strategies
circumvent this bottleneck by selecting special initial states or
engineering the dissipator.
Here we investigate this problem when the input
and the subsequent dissipative generator are prescribed.
We introduce a predissipative geometric steering protocol that
reshapes a prescribed pure or mixed state before
relaxation begins---coherent rotations interleaved with nonselective
projective measurements---at a fixed Lindblad generator.
By steering the state's Bloch direction along geodesic paths,
the protocol suppresses its overlap with the slowest Liouvillian
modes. The prepared state can then start farther from equilibrium yet relax faster, realizing the quantum Mpemba effect, as diagnosed by reduced slow-mode overlap and a larger initial distance to stationarity, with distance crossings for complex slow modes verified from the full dynamics.
Our framework treats real and complex spectral gaps
within a common mode-overlap description.
We also quantify the state-side cost of this
preparation through cumulative measurement-induced mixing and
illustrate its tradeoff with fidelity for a mixed target.
We demonstrate Mpemba acceleration in driven
qubit and multiqubit systems
and discuss the required operations in
trapped-ion and superconducting platforms.
\end{abstract}

\maketitle
\begingroup
\renewcommand{\thefootnote}{}
\endgroup
\section{Introduction}

Understanding how open quantum systems approach their stationary states is
a central problem in nonequilibrium physics~\cite{Dan19,Bea26,Koc22,Gal20,Mei22,Lu17,Ren24,Lan22,Zha26,Cha26Thermo}.
For Markovian dynamics~\cite{Gar04,Bau08}, the Lindblad spectrum sets the
decay timescales, while the late-time pathway also depends on the overlap
of the initial state with the dynamical modes~\cite{Car21,Mor24, Med25,Sha24,Deg22}.

This mode-overlap picture provides a natural explanation of the quantum
Mpemba effect~\cite{Na19,Str25,Vu25,Bao25,Tak21,Lin22,Kli19,Jos24,Liu24,Bia24,Liu24b,Are25}:
a state farther from the stationary state can relax faster than a closer one when it has a smaller overlap with the slowest decaying mode. This is the quantum counterpart of the classical Mpemba effect, originally observed in thermal relaxation~\cite{Mpe69,Kel69,Tez26}. Previous studies have mainly realized the effect by selecting special initial states or special classes of generators~\cite{Med25,Car21,Mor24,Sha24,Bao25,Cha24,Sap26,Are23,Ryl24,Yam26}.
For example, slow-mode suppression can be achieved for pure states with
real spectral gaps by suitable unitary rotations~\cite{Car21}, while
Davies maps provide a structured setting in which quantum detailed
balance strongly constrains the mode decomposition~\cite{Mor24}.

Here we take a different route: a geometric mode-steering protocol prepares
a given pure or mixed state before dissipative relaxation begins. The
preparation stage interleaves coherent evolution with nonselective
projective measurements~\cite{Bra95,Li13,Lis22,Ere08}.
Nonselective measurements have long been recognized as a state-steering tool
in closed systems, where equal angular steps along the connecting geodesic
are optimal for a qubit~\cite{Pec06}. {Pechen \textit{et al.}
also included fixed Hamiltonian evolution between measurements~\cite{Pec06},
and subsequent work developed measurement-assisted control and
steering~\cite{Shu08,Roy20}.} Here this geometric control primitive is embedded
in a different problem: reshaping the overlap of a given state with the
eigenmodes of a fixed dissipative generator.
In the generalized Bloch representation, coherent segments rotate the state
direction, while measurements project it onto the subspaces selected by
their bases. The protocol
therefore steers the state toward directions with reduced slow-mode overlap,
while the subsequent dissipative generator is unchanged.

{The protocol targets settings with a prescribed input and dissipative
generator, where fixed-Hamiltonian evolution and adjustable measurement
bases are available but control during relaxation is restricted.
Preparation is assumed to be effectively isolated and requires no
feedback. We quantify cumulative measurement-induced mixing and its
tradeoff with target-state fidelity. The work, heat, and reset costs of
the apparatus depend on its physical realization~\cite{Kam16,Sag09,Man13};
their joint optimization with relaxation speed is beyond our scope.
Comparisons with unitary preparation, temporary reset, and continuous
control must account for available controls and preparation overheads
as well as relaxation time~\cite{Liu25,Cam17,Bra20,Li22}.
Faster post-preparation relaxation alone implies neither lower total
cost nor a shorter preparation-plus-relaxation time.}

Unlike quantum Zeno and anti-Zeno
protocols~\cite{Mis77,Ita90,Har17,Kof00,Fis01},
in which measurements act during the decay to modify relaxation rates, here
repeated measurements act before dissipation as a preparatory geometric
steering tool. We formulate the idea in a generalized Bloch representation
valid for arbitrary \(d\)-level systems and general Hamiltonians, and derive
how the measurement sequence reshapes the state and its overlap with the
slow Liouvillian modes. The framework encompasses real and complex slow spectral
gaps and is illustrated with a driven dissipative qubit
and coupled multiqubit systems. The required ingredients---coherent control,
repeated nonselective measurements, and tunable Markovian dissipation---are
available in trapped-ion and superconducting-qubit
platforms~\cite{Sha24,Lis22,Koh23}.

\section{Geometric mode-steering framework}
\label{sec:framework}

We consider a $d$-level quantum system in the generalized Bloch
representation~\cite{Byr03,Nie10,Ber08},
\begin{equation}
\rho = \frac{1}{d} I_d + \frac{1}{2} \mathbf{r} \cdot \boldsymbol{\lambda},
\label{rho_gene1}
\end{equation}
where $\mathrm{Tr}(\lambda_i \lambda_j)=2\delta_{ij}$ and
$|\mathbf{r}|\le R_d\equiv \sqrt{2(d-1)/d}$. For $d>2$, this norm bound is necessary but not sufficient for positivity; the states considered below are generated by physical quantum channels. The Hamiltonian is
decomposed as $H=\mathrm{Tr}(H)I_d/d+\mathbf{h}\cdot\boldsymbol{\lambda}/2$;
the dynamics are determined by the traceless part
$H_{\mathrm{eff}}=\mathbf{h}\cdot\boldsymbol{\lambda}/2$, whose unitary
evolution rotates the Bloch vector: $\mathbf{r}(t)=R(t)\mathbf{r}(0)$.

We implement the geometric mode-steering stage by \(N+1\) coherent
unitary segments interleaved with \(N\) nonselective projective
measurements. The unitary segments are
$U_k = e^{-i H_{\mathrm{eff}} \Delta t_k}$, \(k=1,\ldots,N+1\), with total
coherent-evolution duration \(T=\sum_{k=1}^{N+1}\Delta t_k\).
The ideal measurement maps are instantaneous; physical measurement and
basis-control durations must be included in the full preparation time.
Each measurement 
\( 
\tilde{\mathcal{M}}_{\mathbf{a}_k}[\rho] = \sum_\mu \tilde{\Pi}_{\mathbf{a}_k}^{(\mu)} \rho \, \tilde{\Pi}_{\mathbf{a}_k}^{(\mu)}
\)
acts through orthogonal projectors. Denoting the cumulative unitary before the $k$th measurement as $V_k = U_k U_{k-1} \cdots U_1$, the rotated measurement satisfies
\( 
\tilde{\mathcal{M}}_{\mathbf{a}_k}[V_k \rho V_k^\dagger] = V_k \mathcal{M}_{\mathbf{a}_k}[\rho] V_k^\dagger,
\) 
where $\mathcal{M}_{\mathbf{a}_k} = V_k^\dagger \tilde{\mathcal{M}}_{\mathbf{a}_k} V_k$ denotes the measurement in the rotating frame in superoperator notation, with projectors $\Pi_{\mathbf a_k}^{(\mu)}=V_k^\dagger\tilde\Pi_{\mathbf a_k}^{(\mu)}V_k$. Collecting all unitary controls into the final rotation $U_{\rm tot} = U_{N+1} U_N \cdots U_1 = e^{-i H_{\rm eff} T}$, the prepared state becomes
\( 
\tilde{\rho}_N = U_{\rm tot} \rho_N U_{\rm tot}^\dagger, \) where \( 
\rho_N = \mathcal{M}_{\mathbf{a}_N} \circ \cdots \circ \mathcal{M}_{\mathbf{a}_1}[\rho_0].
\)
This completes the  steering stage. 

In the Bloch representation, each unitary induces a rotation $\mathbf{r} \mapsto R_k \mathbf{r}$, while each measurement projects as
\( 
\mathcal{M}_{\mathbf{a}_k}[\rho] = \frac{I_d}{d} + (\mathcal{P}_{\mathbf{a}_k} \mathbf{r}) \cdot \boldsymbol{\lambda}/2.
\)
Here $\mathcal P_{\mathbf a_k}$ projects onto the traceless operator subspace left invariant by the measurement; for a complete rank-1 measurement it has dimension $d-1$, reducing to a Bloch axis for a qubit. Thus, the Bloch vector of the prepared state is
\begin{equation}
\tilde{\mathbf{r}}_N = |\mathbf{r}_0| \, R_{\rm tot} \, \mathcal{T}_N \, \hat{\mathbf{n}}_0,
\label{rnn0}
\end{equation}
with \(R_{\rm tot} = \prod_{k=1}^{N+1} R_k\),
\(\mathcal{T}_N = \prod_{k=1}^N \mathcal{P}_{\mathbf{a}_k}\),
with the earliest operation acting on the right, and
\(\hat{\mathbf{n}}_0 = \mathbf{r}_0 / |\mathbf{r}_0|\)
for $|\mathbf r_0|>0$. The maximally mixed input is unchanged.

We select a target state
$|\tau\rangle$ whose overlap with the slowest left eigenoperator is
small while its distance from the stationary state is large; the
control stage is then designed to maximize the target population
$\tilde{p}_{\tau}=\langle\tau|\tilde{\rho}_N|\tau\rangle$. Writing
$|\tau\rangle\langle\tau|=I_d/d+\mathbf{r}_{\tau}\cdot\bm{\lambda}/2$
with $|\mathbf{r}_{\tau}|=R_d$, the generalized Bloch representation
yields
\begin{equation}
\tilde{p}_{\tau}=\frac{1}{d}+\frac{1}{2}|\mathbf{r}_0|R_d\,
\hat{\mathbf{r}}_{\tau}\cdot\bigl(R_{\rm tot}\mathcal{T}_N\hat{\mathbf{n}}_0\bigr).
\label{eq:target_population}
\end{equation}

{For the pure target
$\rho_\tau=|\tau\rangle\langle\tau|$,
the target population coincides with the squared Uhlmann fidelity
$F(\tilde\rho_N,\rho_\tau)
=
\left[
\mathrm{Tr}
\sqrt{
\sqrt{\rho_\tau}\tilde\rho_N\sqrt{\rho_\tau}
}
\right]^2
=
\tilde p_\tau $.
The construction therefore accommodates arbitrary pure or mixed input
states while retaining a pure target. Unitality ensures
$\tilde p_\tau\le\lambda_{\max}(\rho_0)$.
For mixed targets, fidelity must instead be evaluated from the full
density matrices.}

\begin{figure}[t]
\centering
\hspace*{0.03\columnwidth}
\resizebox{0.85\columnwidth}{!}{%
\begin{tikzpicture}[
    x=1cm,
    y=1cm,
    >=latex,
    font=\small,
    thickarrow/.style={->, line width=1.2pt},
    unitary/.style={
        draw=blue!80!black,
        fill=blue!85!black,
        text=white,
        line width=0.5pt,
        minimum width=0.95cm,
        minimum height=0.75cm,
        align=center
    },
    measure/.style={
        draw=red!80!black,
        fill=red!80!black,
        text=white,
        line width=0.5pt,
        minimum width=1.05cm,
        minimum height=0.75cm,
        align=center
    }
]
\node at (-4.625,1.50) {\Large $\rho_0$};

\begin{scope}[shift={(-5.75,1.10)}, scale=0.5625]
    \fill[red!65!yellow] (0,0) rectangle (4,-4);
    \fill[blue!80] (0,0) rectangle (1,-1);
    \fill[blue!80] (1,-1) rectangle (2,-2);
    \fill[blue!80] (2,-2) rectangle (3,-3);
    \fill[blue!80] (3,-3) rectangle (4,-4);
    \draw[line width=0.55pt] (0,0) rectangle (4,-4);
    \foreach \x in {1,2,3}{
        \draw[line width=0.55pt] (\x,0) -- (\x,-4);
    }
    \foreach \y in {-1,-2,-3}{
        \draw[line width=0.55pt] (0,\y) -- (4,\y);
    }
\end{scope}
\draw[thickarrow] (-3.30,0.00) -- (-2.82,0.00);
\draw[
    black,
    rounded corners=7pt,
    line width=0.8pt
] (-2.75,-1.95) rectangle (3.35,2.00);

\node at (0.30,1.45) {\Large Isolated steering protocol};
\node[unitary] at (-2.10,0.30) {\normalsize $U_1$};
\node[measure] at (-0.80,0.30) {$\tilde{\mathcal M}_{\mathbf{a}_1}$};

\node at (0.25,0.30) {\Large $\cdots$};

\node[measure] at (1.30,0.30) {$\tilde{\mathcal M}_{\mathbf{a}_N}$};
\node[unitary] at (2.6,0.30) {\normalsize $U_{N+1}$};
\draw[thickarrow] (-2.25,-0.65) -- (3.05,-0.65);

\node at (0.35,-1.20) {\Large $\mathbf r \mapsto R_{\rm tot} \mathcal{T}_N\mathbf r$};
\draw[thickarrow] (3.45,0.00) -- (4.00,0.00);
\node at (5.225,1.50) {\Large $\tilde{\rho}_N$};

\begin{scope}[shift={(4.10,1.10)}, scale=0.5625]
    \fill[white] (0,0) rectangle (4,-4);
    \fill[blue!100] (0,0) rectangle (1,-1);
    \fill[blue!60] (1,-1) rectangle (2,-2);
    \fill[blue!60] (2,-2) rectangle (3,-3);
    \fill[blue!60] (3,-3) rectangle (4,-4);
    \draw[line width=0.55pt] (0,0) rectangle (4,-4);
    \foreach \x in {1,2,3}{
        \draw[line width=0.55pt] (\x,0) -- (\x,-4);
    }
    \foreach \y in {-1,-2,-3}{
        \draw[line width=0.55pt] (0,\y) -- (4,\y);
    }
    \node[text=white, font=\small] at (0.50,-0.50) {$|\tau\rangle$};
\end{scope}
\begin{scope}[shift={(-5.75,-2.55)}, scale=0.5625]
    \fill[blue!70] (0,0) rectangle (0.75,-0.75);
    \draw[line width=0.55pt] (0,0) rectangle (0.75,-0.75);
\end{scope}
\node[anchor=west] at (-5.20,-2.76) {\large population};

\begin{scope}[shift={(-2.65,-2.55)}, scale=0.5625]
    \fill[red!65!yellow] (0,0) rectangle (0.75,-0.75);
    \draw[line width=0.55pt] (0,0) rectangle (0.75,-0.75);
\end{scope}
\node[anchor=west] at (-2.10,-2.76) {\large coherence};
\node at (6.30,2.05) {\Large (a)};

\end{tikzpicture}%
}
\par\vspace{0.22cm}
\begin{minipage}{0.49\columnwidth}
\centering
\begin{overpic}[width=3.4cm]{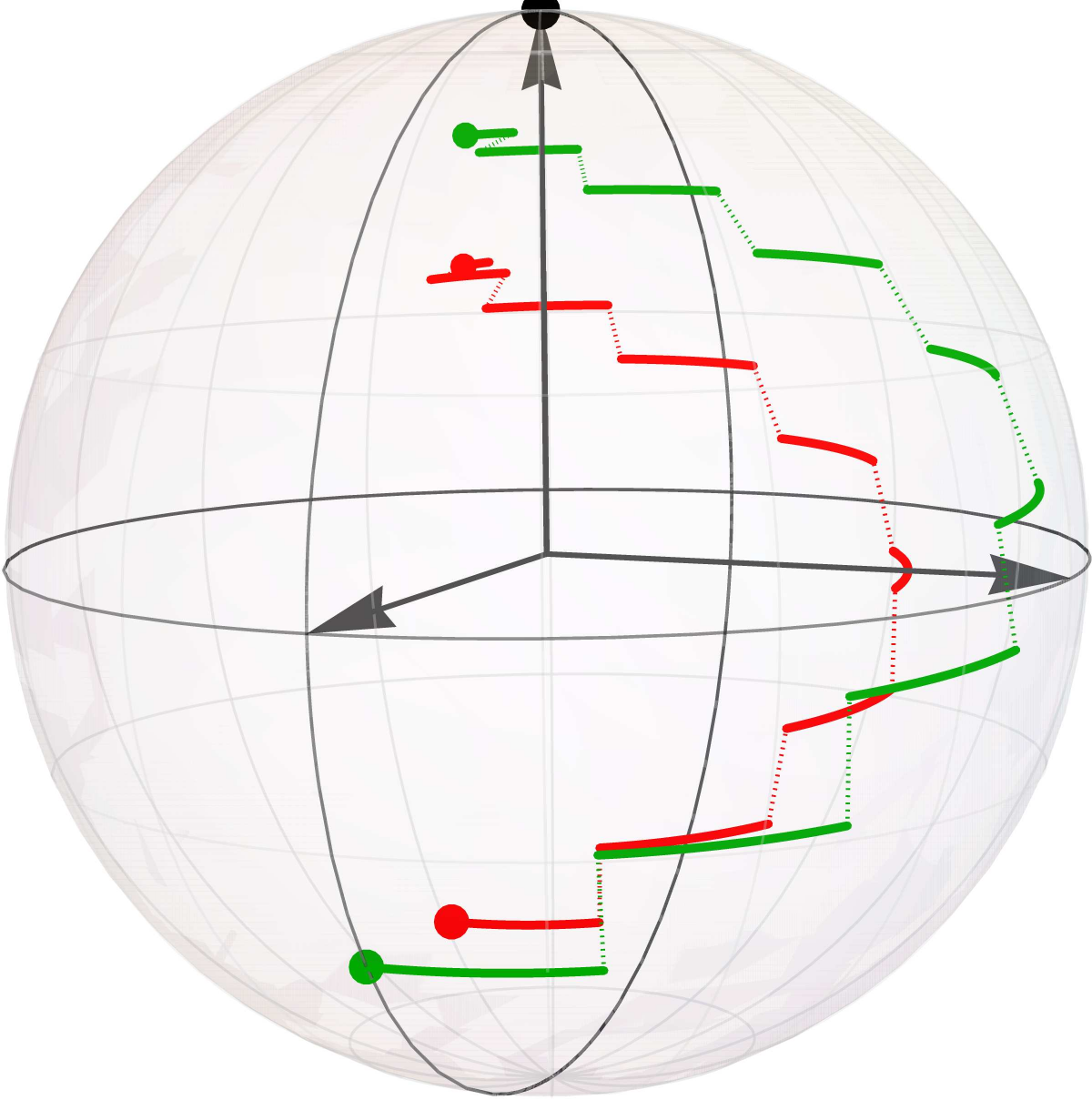}
\put(85,101){\scalebox{0.9}{(b)}}
\put(31,5){\scalebox{0.8}{$\bm{\rho_0}$}}
\put(40,22){\scalebox{0.8}{$\bm{\rho_0'}$}}
\put(21,38){\scalebox{0.8}{$r_x$}}
\put(99,45){\scalebox{0.8}{$r_y$}}
\put(48,102){\scalebox{0.8}{$r_z$}}
\put(36,92){\scalebox{0.8}{$\bm{\tilde{\rho}_N}$}}
\put(36,66){\scalebox{0.8}{$\bm{\tilde{\rho}_N'}$}}
\end{overpic}
\end{minipage}
\hfill
\begin{minipage}{0.49\columnwidth}
\centering
\begin{overpic}[width=3.5cm]{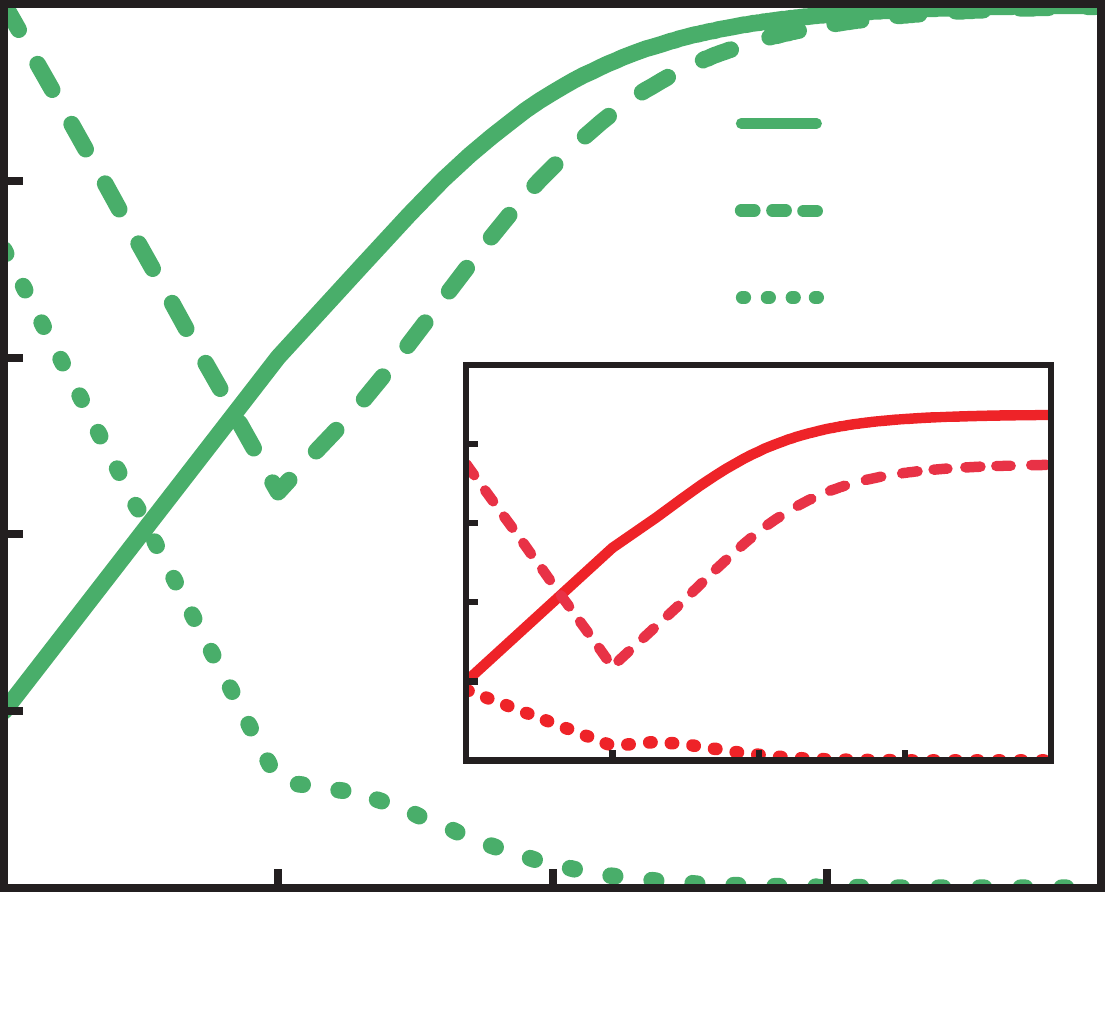}
\put(82,98){\scalebox{0.9}{(c)}}
\put(-5,90){\scalebox{0.8}{1}}
\put(-11,75){\scalebox{0.8}{0.8}}
\put(-11,59){\scalebox{0.8}{0.6}}
\put(-11,43){\scalebox{0.8}{0.4}}
\put(-11,27){\scalebox{0.8}{0.2}}
\put(-5,11){\scalebox{0.8}{0}}
\put(-1,5){\scalebox{0.8}{0}}
\put(23,5){\scalebox{0.8}{$1$}}
\put(45,5){\scalebox{0.8}{$10$}}
\put(69,5){\scalebox{0.8}{$100$}}
\put(90,5){\scalebox{0.8}{$1000$}}
\put(37,24){\scalebox{0.6}{0}}
\put(37,57.5){\scalebox{0.6}{1}}
\put(40,19.5){\scalebox{0.6}{0}}
\put(87,19){\scalebox{0.6}{$1000$}}
\put(45,-3.5){\scalebox{1.0}{$\bm{N}$}}
\put(77,81){\scalebox{0.7}{$\tilde{p}_{e}^\text{max} $}}
\put(77,72.5){\scalebox{0.7}{$|\tilde{\mathbf{r}}_N|$}}
\put(77,64.5){\scalebox{0.6}{$\mathcal{C}(\tilde\rho_N)$}}
\end{overpic}
\end{minipage}

\caption{(a) Geometric mode-steering protocol. The sequence of \(N+1\)
unitary segments and \(N\) nonselective measurements maps
\(\mathbf r \mapsto R_{\rm tot}\mathcal T_N \mathbf r\), {producing the prepared state} \(\tilde\rho_N\).
(b) Bloch trajectories for $N=8$ measurements, steering the pure
$\rho_0$ \((|\mathbf r_0|=1,\theta=\arccos(-0.6), \phi=0\)) and mixed $\rho_0'$ \((|\mathbf r_0'|=0.75,\theta=\arccos(-0.8), \phi=0)\) initial states
to \(\tilde\rho_N\) and \(\tilde\rho_N'\), respectively. Solid arcs:
unitaries; dotted chords: measurement projections.
(c) Optimal population \(\tilde p_e^{\,\max}\), Bloch length
\(|\tilde{\mathbf r}_N|\), and coherence \(\mathcal C(\tilde\rho_N)\) vs
\(N\); inset: mixed input. The solid curve is bounded by
\((1+|\mathbf r_0|)/2\). Parameters: \(T=0.1\), \(\Omega=0.1\),
\(\Delta=2.0\).}
\label{combined}
\end{figure}

We refer to sequences for which
\(R_{\rm tot}\mathcal{T}_N\hat{\bf n}_0\) is aligned with
\(\hat{\bf r}_\tau\) as target-aligned steering sequences. At fixed output Bloch length, this alignment
maximizes the target population \(\tilde p_\tau\): coherent segments rotate
the state direction, while nonselective measurements suppress transverse
components relative to the chosen measurement subspaces. The construction uses this geometry to enhance the target population
[Fig.~\ref{combined}(a)]. For a general mixed multilevel input,
the remaining state components must also be retained when evaluating
the slow-mode overlaps.
{For qubits, the measurement bases depend on the
input Bloch direction, while purity limits the attainable target
population. Multiqubit implementations also require a specified
steering subspace and realizable basis changes; these information
and control requirements are compared with those of unitary
preparation~\cite{Koc22,Car21,Mor24} in
Appendix~\ref{app:comparison}.}

\section{Liouvillian-mode suppression and Mpemba conditions}
\label{sec:mode_conditions}
After the steering stage, the measurement sequence is stopped and the
prepared state \(\tilde\rho_N\) is used as the initial state for relaxation
under a fixed Markovian generator~\cite{Bre02,Lin76},
\begin{equation}
\dot{\rho}_t = \mathcal{L}[\rho_t]
= -i[H,\rho_t]+\sum_m\Bigl(L_m\rho_t L_m^\dagger
-\tfrac{1}{2}\{L_m^\dagger L_m,\rho_t\}\Bigr),
\label{lrhol}
\end{equation}
with a unique steady state $\rho_{\mathrm{ss}}$ and spectral decomposition
$\rho_t=\rho_{\mathrm{ss}}+\sum_{\nu\ge2}\mathrm{Tr}(l_\nu\rho_0)r_\nu e^{\lambda_\nu t}$.
This spectral expansion assumes a diagonalizable generator. We use
trace-dual left operators, with $\mathcal L^\dagger[l_\nu]=\lambda_\nu l_\nu$
and $\mathrm{Tr}(l_\nu r_\mu)=\delta_{\nu\mu}$; the coefficient is
$\mathrm{Tr}(l_\nu\rho)$ without a dagger. For an isolated real slow
mode, the leading long-time contribution is:
$\rho_t-\rho_{\mathrm{ss}}\simeq\mathrm{Tr}(l_2\rho_0)r_2 e^{\lambda_2 t}$.

Expanding the left eigenoperator as $l_2=\beta_0 I_d+\bm{\beta}\cdot\boldsymbol{\lambda}$,
\begin{equation}
\mathrm{Tr}(l_2\tilde{\rho}_N)=\beta_0+\bm{\beta}\cdot(R_{\mathrm{tot}}\mathbf{r}_N),
\label{trrn}
\end{equation}
from which the slow-mode amplitude ratio is
\begin{equation}
\mathcal{R}=\frac{\mathrm{Tr}(l_2\tilde{\rho}_N)}{\mathrm{Tr}(l_2\rho_0)}
=\frac{\beta_0+|\mathbf{r}_0|\bm{\beta}\cdot(R_{\mathrm{tot}}\mathcal{T}_N\hat{\mathbf{n}}_0)}
{\beta_0+\bm{\beta}\cdot\mathbf{r}_0}.
\label{rr0}
\end{equation}

The condition \(|\mathcal R|<1\) means the prepared state carries a smaller slow-mode amplitude and, for an isolated real slow mode, therefore relaxes faster at long times. When \(|\mathcal R|\to0\), the slowest
mode is effectively removed. The ratio is understood for \(\mathrm{Tr}(l_2\rho_0)\neq0\); a vanishing denominator means the initial state already has no slow-mode overlap.

Equation~\eqref{rr0} applies irrespective of whether the slowest nonzero eigenvalue is real or belongs to a complex-conjugate pair. For a
real \(\lambda_2\), the long-time contribution is proportional to
\(\mathrm{Tr}(l_2\rho_0)r_2e^{\lambda_2t}\), and suppression of
\(\mathrm{Tr}(l_2\tilde\rho_N)\) directly accelerates relaxation. For a complex pair $(\lambda_2,\lambda_2^*)$, the slow contribution contains both
$l_2$ and $l_2^\dagger$, the latter being the left eigenoperator associated with
$\lambda_2^*$. Since $\tilde\rho_N$ is Hermitian,
$\mathrm{Tr}(l_2^\dagger\tilde\rho_N)=[\mathrm{Tr}(l_2\tilde\rho_N)]^*$, so the two
members of the pair carry identical amplitude moduli and the single ratio
$|\mathcal{R}|$ of Eq.~\eqref{rr0} controls the suppression of both. In both cases the protocol acts by the same geometric mechanism: steering \(R_{\rm tot}\mathcal T_N\hat{\mathbf n}_0\) toward directions of reduced slow-mode overlap.

The quantum Mpemba effect requires two conditions: the prepared state
must have a smaller slow-mode amplitude and must initially be farther
from the stationary state. We quantify the latter by the trace
distance~\cite{Car21,Nie10,Man21}, which, unlike the nonequilibrium free
energy used for Davies maps~\cite{Mor24}, remains well defined for
generic Liouvillians without a thermal fixed point:
\begin{equation}
D_T(\rho, \rho_{\mathrm{ss}})
=
\frac{1}{2}\lVert \rho-\rho_{\mathrm{ss}}\rVert _1 .
\end{equation}
Accordingly, the candidate Mpemba regime is identified by
\(|\mathcal R|<1\) and \(\kappa_0>1\),  where
\(
\kappa_0\equiv 
{D_T(\tilde{\rho}_N, \rho_{\mathrm{ss}})}/
     {D_T(\rho_0, \rho_{\mathrm{ss}})}.
\) For complex slow modes, \(|\mathcal R|<1\) reduces the long-time envelope; distance crossings are verified from the full dynamics. This choice is further justified in Appendix~\ref{sec:trace_vs_relative}, where coherences are shown to
contribute at the same linear order as populations in the qubit example. When
\(|\mathcal R|\to0\) and \(\kappa_0>1\), the protocol realizes a strong
quantum Mpemba effect.

We illustrate this framework for both pure and mixed initial states;
no quantum detailed balance is assumed, in contrast to the Davies-map
setting~\cite{Mor24,Dav74,Bre02}.
A driven dissipative qubit demonstrates the protocol beyond the Davies-map
limit, including real and complex slow spectral gaps. A coupled multiqubit
system then shows that steering the state toward the fully excited product
state strongly suppresses coherence-dominated slow modes. In the coherence-sector benchmark, the slow-mode ratio approaches
\( |\mathcal R|\to0 \) when the prepared state becomes orthogonal to
all relevant slow coherence modes; a state diagonal in the corresponding
basis provides this cancellation.

\section{Applications to driven dissipative systems}
\label{sec:applications}

\subsection{Driven dissipative qubit}
\label{sec:qubit}
\label{subsec:example_qubit}

\subsubsection{State preparation}
\label{subsubsec:qubit_preparation}
We first consider a driven dissipative qubit with Hamiltonian
\begin{equation}
H=\Delta |e\rangle\langle e|
+\frac{\Omega}{2}(\sigma^++\sigma^-),
\label{eq:qubit_H}
\end{equation}
where \(\Delta\) is the detuning and \(\Omega\) is the coherent coupling
strength~\cite{Zha22}.  For this two-level example the target is the excited state,
$|\tau\rangle=|e\rangle$, so that $\tilde{p}_{\tau}$ reduces to the
excited-state population $\tilde{p}_e$. With driving and detuning present, the model leaves the Davies-map setting and provides a minimal test of the protocol beyond detailed-balance dynamics. A single-qubit Davies map can have a complex spectral gap~\cite{Mor24}; with driving, the gap can be tuned between real and complex, realizing both cases of Eq.~\eqref{rr0} in one model.

{
Writing the initial Bloch vector as
$\mathbf r_0
=
|\mathbf r_0|
(\sin\theta\cos\phi,
\sin\theta\sin\phi,
\cos\theta)$,
the steering stage admits a simple geometric description, as illustrated
in Fig.~\ref{combined}(b). For a sequence of \(N\) nonselective
measurements interleaved with \(N+1\) unitary segments, it is convenient
to work in the rotating frame defined by the cumulative unitary
evolution. Denoting the effective measurement axes by
\(\mathbf a_k\), each nonselective measurement removes the component of
the Bloch vector perpendicular to \(\mathbf a_k\),
\begin{equation}
\mathbf r_k
=
(\mathbf a_k\cdot\mathbf r_{k-1})\mathbf a_k
=
|\mathbf r_{k-1}|
\cos\theta_k\,\mathbf a_k ,
\label{eq:measurement_projection}
\end{equation}
where \(\theta_k\) is the angular separation between two consecutive
effective directions.  After \(N\) measurements,
\begin{equation}
\mathbf{r}_N = |\mathbf{r}_0| \prod_{k=1}^{N} \cos \theta_k \, \mathbf{a}_{N}.
\label{eq:qubit_RN}
\end{equation}
This expression displays the multiplicative contraction induced by repeated nonselective measurements.

The coherent rotations can be collected into the total unitary
\(U_{\rm tot}\), so that the physical target \(|e\rangle\) is mapped in
the rotating frame to the effective target direction
$|e_{\rm eff}\rangle
=
U_{\rm tot}^{\dagger}|e\rangle$.
Equivalently, at the Bloch-vector level the target axis is
\(\mathbf a_{\rm eff}=R_{\rm tot}^{-1}\hat{\mathbf z}\), where \(\hat{\mathbf z}\) is the Bloch-sphere direction corresponding to the excited state \(|e\rangle\).
This effective target axis can be regarded as the \((N+1)\)-th
measurement axis, so that the final projection of the Bloch vector onto
the physical target direction is
\begin{align}
\tilde r_z
&=
\tilde{\mathbf r}_N\cdot\hat{\mathbf z}
\nonumber\\[2pt]
&=
|\mathbf r_0|
\prod_{k=1}^{N}\cos\theta_k
\cos\theta_{N+1},
\label{eq:rz}
\end{align}
where we have used
$\tilde{\mathbf r}_N\cdot\hat{\mathbf z}
=
\mathbf r_N\cdot\mathbf a_{\rm eff}$,
and $\theta_{N+1}$ denotes the angular separation between the last
measurement axis $\mathbf a_N$ and the effective target direction
$\mathbf a_{\rm eff}$.
Including the final angular step from the last measurement axis to
\(\mathbf a_{\rm eff}\), the prepared excited-state population is
\begin{align}
\tilde{p}_e
&=
\frac{1}{2}
+
\frac{1}{2}
(\tilde{\mathbf r}_N \cdot \hat{\mathbf z})
\nonumber\\[2pt]
&=
\frac{1}{2}
+
\frac{1}{2}
|\mathbf r_0|
\prod_{k=1}^{N+1}\cos\theta_k,
\label{eq:SM_pe}
\end{align}

Within this geodesic class, the total angular distance is
fixed by
$\sum_{k=1}^{N+1}\theta_k
=
\delta\theta_{\rm eff}$,
where \(\delta\theta_{\rm eff}\) is the geodesic angle between the
initial Bloch direction
\(\hat{\mathbf n}_0=\mathbf r_0/|\mathbf r_0|\)
and the effective target direction.

Under this constraint, maximizing the target population therefore
amounts to maximizing
\(\prod_{k=1}^{N+1}\cos\theta_k\)
under the constraint $\sum_{k=1}^{N+1}\theta_k
=
\delta\theta_{\rm eff}$. For $0\le\theta_k<\pi/2$, strict concavity of $\ln(\cos\theta)$
gives
\begin{equation}
\sum_{k=1}^{N+1}\ln\cos\theta_k
\le
(N+1)
\ln\cos\left(
\frac{\delta\theta_{\rm eff}}{N+1}
\right),
\label{eq:jensen_geodesic}
\end{equation}
with equality if and only if all angular increments are equal. The
optimal geodesic schedule is therefore
\begin{equation}
\theta_k
=
\frac{\delta\theta_{\rm eff}}{N+1},
\qquad
k=1,\ldots,N+1 .
\label{eq:uniform_geodesic}
\end{equation}
A sequence containing an increment at least $\pi/2$ cannot
improve the nonnegative equal-step product; the boundary case
$N=1$, $\delta\theta_{\rm eff}=\pi$ follows by continuity.
This reproduces the equal-step optimality of measurement-driven state
steering in closed systems~\cite{Pec06}, while here the effective target
is fixed by the coherent evolution and the Liouvillian spectral
structure rather than by an arbitrary state-transfer objective. The
result establishes optimality within the geodesic class at fixed
\(\delta\theta_{\rm eff}\); global optimality over arbitrary
non-geodesic measurement sequences is not assumed.

Substituting Eq.~\eqref{eq:uniform_geodesic} into
Eq.~\eqref{eq:SM_pe} yields
\begin{equation}
\tilde p_e^{\rm max}
=
\frac{1}{2}
+
\frac{1}{2}
|\mathbf r_0|
\left[
\cos\left(
\frac{\delta\theta_{\rm eff}}{N+1}
\right)
\right]^{N+1}.
\label{eq:pe_max}
\end{equation}
At finite $N$, the full prepared vector is
\begin{equation}
\tilde{\mathbf r}_N=|\mathbf r_0|
\cos^N\!\left(\frac{\delta\theta_{\rm eff}}{N+1}\right)
R_{\rm tot}\mathbf a_N,
\label{eq:finiteN_qubit}
\end{equation}
where $\mathbf a_N$ lies a fraction $N/(N+1)$ along the geodesic.
Its residual tilt and contraction must both be retained when evaluating
finite-$N$ modal overlaps.
For large \(N\),
\begin{equation}
\left[
\cos\left(
\frac{\delta\theta_{\rm eff}}{N+1}
\right)
\right]^{N+1}
\simeq
\exp\left[
-\frac{\delta\theta_{\rm eff}^{2}}
{2(N+1)}
\right],
\label{eq:geometric_scaling}
\end{equation}
so that the finite-measurement geometric loss,
\begin{equation}
\epsilon_{\rm geo}
\equiv
1-
\left[
\cos\left(
\frac{\delta\theta_{\rm eff}}{N+1}
\right)
\right]^{N+1},
\label{eq:geometric}
\end{equation}
obeys
$\epsilon_{\rm geo}
\simeq
\frac{\delta\theta_{\rm eff}^{2}}
{2(N+1)}$.
The steering error therefore decreases algebraically as \(1/N\), while
the required number of measurements grows correspondingly.

The equal-step geodesic schedule is determined by the Bloch-vector
direction and therefore applies to both pure and mixed initial states
with \(|\mathbf r_0|>0\). The Bloch-vector length instead sets the
achievable target population through the prefactor
\(|\mathbf r_0|\) in Eq.~\eqref{eq:pe_max}. Accordingly,
\(\tilde p_e^{\rm max}\to1\) for a pure initial state as
\(N\to\infty\), whereas for a mixed state it approaches
\((1+|\mathbf r_0|)/2\). The maximally mixed state
\((|\mathbf r_0|=0)\) has no well-defined Bloch direction and is
therefore excluded from this directional steering construction.}

Figure~\ref{combined}(c) confirms the effectiveness of the control stage. The maximal population
$\tilde p_e^{\max}$ increases monotonically with $N$, approaching unity for
pure initial states and saturating below unity for mixed states because the
initial Bloch vector is shorter. The Bloch-vector length $|\tilde{\bf r}_N|$
quantifies this geometric cost: a single large-angle projection ($N=1$)
sharply contracts the state, after which $|\tilde{\bf r}_N|$ recovers toward
$|{\bf r}_0|$ as the per-step angle $\delta\theta_{\rm eff}/(N+1)$ shrinks. The remaining coherence is measured in the computational basis by
\(\mathcal C(\rho)=S[\Lambda_{\rm com}(\rho)]-S(\rho)\)~\cite{Bau14,Str17,Xia23},
where \(S(\rho)=-\mathrm{Tr}(\rho\ln\rho)\) and \(\Lambda_{\rm com}\) is the
complete-dephasing map. The decrease of $\mathcal{C}(\tilde\rho_N)$ with $N$ reflects the
progressive suppression of coherence components that may overlap with
slow Liouvillian modes and can thereby reduce the contribution of the slow
relaxation mode. For a driven Liouvillian this connection requires an
explicit overlap calculation; computational-basis coherence alone does
not determine the slow-mode weight.

\subsubsection{Relaxation dynamics}
\label{subsubsec:qubit_relaxation}
\begin{figure}[!t]
\centering
\setlength{\textfloatsep}{0pt}
\setlength{\abovecaptionskip}{15pt}   
\includegraphics[width=7cm]{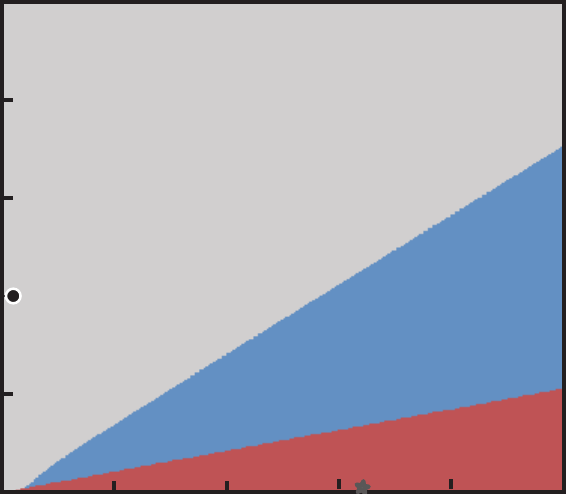}
\put(-110,5){\scalebox{0.9}{\color{white}\shortstack{Real gap\\ (both branches required)}}}
\put(-120,33){\scalebox{0.9}{\rotatebox{20}{\color{white}\shortstack{Real gap\\ (linear reduction valid)}}}}
\put(-165,120){\scalebox{0.9}{\shortstack{Complex gap\\ (modulus condition)}}}
\put(-220,80){\rotatebox{90}{\scalebox{1.1}{$\bm{\Delta}$}}}
\put(-100,-20){\scalebox{1.1}{$\bm{\Omega}$}}
\put(-210,170){\scalebox{1.1}{5}}
\put(-210,135){\scalebox{1.1}{4}}
\put(-210,100){\scalebox{1.1}{3}}
\put(-210,65){\scalebox{1.1}{2}}
\put(-210,31){\scalebox{1.1}{1}}
\put(-210,0){\scalebox{1.1}{0}}
\put(-200,-10){\scalebox{1.1}{0}}
\put(-162,-10){\scalebox{1.1}{$1$}}
\put(-123,-10){\scalebox{1.1}{$2$}}
\put(-82,-10){\scalebox{1.1}{$3$}}
\put(-43,-10){\scalebox{1.1}{$4$}}
\put(-3,-10){\scalebox{1.1}{$5$}}
\caption{Validity domain of the linear reduction \eqref{eq:SM_ratio} in the
$(\Omega,\Delta)$ plane for $\gamma=1$ and $|\mathbf r_0|=1$ (the most
restrictive case). Gray: the slowest nonzero eigenvalue $\lambda_2$
belongs to a complex-conjugate pair, and the modulus condition
\eqref{eq:SM_modulus} with complex coefficients applies.
Blue: real spectral gap satisfying the sufficient condition
\eqref{eq:SM_suff}, where the acceleration condition reduces to the
single branch $X>1$, Eq.~\eqref{eq:SM_ratio}.
Red: real spectral gap where Eq.~\eqref{eq:SM_suff} fails and both
branches of Eq.~\eqref{eq:SM_two_branches} must be retained.
The dot and the star mark {$(\Omega,\Delta)=(0.1,2)$ and $(3.2,0.05)$},
respectively. The boundaries refer to the target-aligned limit.}
\label{fig:branchmap}
\end{figure}

We then let the prepared state relax under the zero-temperature
Lindblad equation
\begin{equation}
\dot{\rho}
=
-i[H,\rho]
+
\frac{\gamma}{2}
\left(
2\sigma^-\rho\sigma^+
-\sigma^+\sigma^-\rho
-\rho\sigma^+\sigma^-
\right),
\end{equation}
where $\gamma$ is the spontaneous-emission rate, $\sigma^+=\ket{e}\bra{g}$ and $\sigma^-=\ket{g}\bra{e}$ are the raising and lowering operators.
{For an initial Bloch direction
$(\theta,\phi)$, with
$\mathbf r_0
=
|\mathbf r_0|
(\sin\theta\cos\phi,
\sin\theta\sin\phi,
\cos\theta)$,
in the target-aligned frequent-measurement limit, the slow-mode amplitude
ratio can be written as
$\mathcal R_\infty
=
(a+1)/(a+X)$,
where
$X=b_x\sin\theta\cos\phi
+b_y\sin\theta\sin\phi
+\cos\theta$.
Here $a=-\gamma/(|\mathbf r_0|\lambda_2)$,
$b_x=4\Omega\Delta/D$, 
$b_y=2\Omega(\gamma+2\lambda_2)/D$, with $D=(\gamma+2\lambda_2)^2+4\Delta^2$ (see Appendix~\ref{app:accel_condition}).
Here $\lambda_2$ denotes the nonzero Liouvillian eigenvalue with the
largest real part.
The slow-mode suppression condition
$|\mathcal R_\infty|<1$ is therefore equivalent to
\begin{equation}
|a+1|<|a+X|.
\label{eq:SM_modulus}
\end{equation}
This condition applies to both real and complex $\lambda_2$ and directly
identifies the initial-state orientations and driving parameters that
suppress the slow Liouvillian mode.

For a complex spectral gap, $a$, $b_x$, $b_y$, and $X$ are generally
complex, and the full modulus condition
\eqref{eq:SM_modulus} must be retained. For a real spectral gap
($\lambda_2<0$), these quantities are real and $a>0$, so that
Eq.~\eqref{eq:SM_modulus} becomes
\begin{equation}
(X-1)(X+1+2a)>0,
\label{eq:SM_two_branches}
\end{equation}
corresponding to the two branches
$X>1$ and $X<-(1+2a)$.
By the Cauchy--Schwarz inequality,
$|X|\le\sqrt{b_x^2+b_y^2+1}$, so the second branch is empty---and the
acceleration condition reduces to a single linear inequality---whenever
\begin{equation}
1+2a
\;\ge\;
\sqrt{b_x^2+b_y^2+1}.
\label{eq:SM_suff}
\end{equation}
Since $a=\gamma/(|\mathbf r_0|\,|\lambda_2|)$ decreases with increasing
$|\mathbf r_0|$, validity of Eq.~\eqref{eq:SM_suff} at
$|\mathbf r_0|=1$ implies its validity for all mixed states.
In this regime, the suppression condition reduces to the linear form
\begin{equation}
1
<
b_x\sin\theta\cos\phi
+
b_y\sin\theta\sin\phi
+
\cos\theta .
\label{eq:SM_ratio}
\end{equation}

For $|\mathbf r_0|=1$, the validity of this linear reduction is mapped
in the $(\Omega,\Delta)$ plane in Fig.~\ref{fig:branchmap}.
The reduction holds over the moderately driven part of the real-gap
region, but fails at strong driving and small detuning, where
$\lambda_2\rightarrow-\gamma/2$ and
$D=(\gamma+2\lambda_2)^2+4\Delta^2$ becomes small, so that both branches
of Eq.~\eqref{eq:SM_two_branches} must be retained.
This distinction concerns only the reduced linear form; the full modulus
condition \eqref{eq:SM_modulus} remains valid wherever its
parametrization is nonsingular in the target-aligned limit and is used
to calculate the displayed amplitude boundaries.}

\begin{figure}[t]
\centering

\makebox[\columnwidth][c]{%
\begin{minipage}[t]{0.34\columnwidth}
\vspace{0pt}
\centering

\begin{overpic}[width=0.9\linewidth]{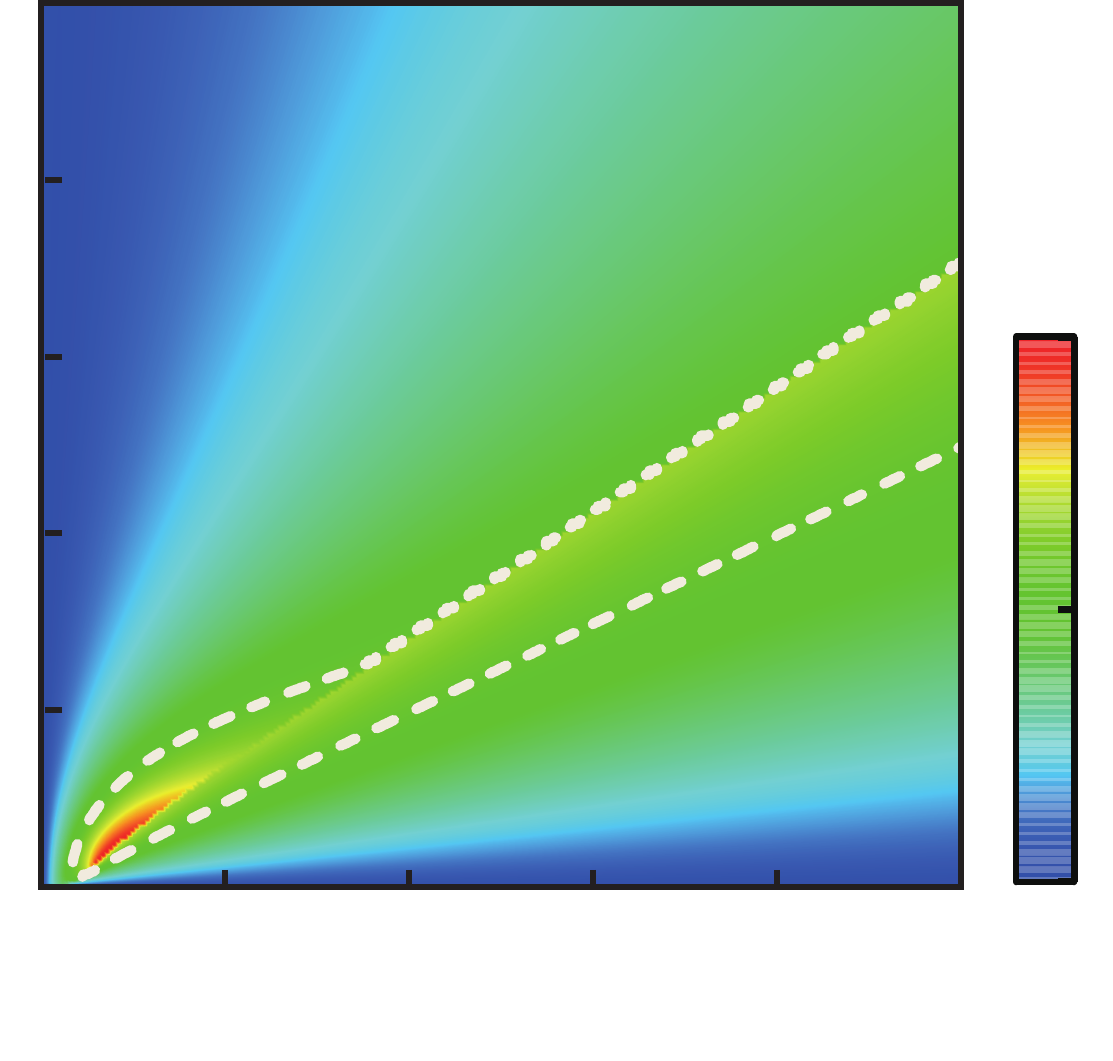}
\put(70,85){\scalebox{0.75}{\textbf{(a)}}}
\put(-15,50){\rotatebox{90}{\scalebox{0.7}{$\bm{\Delta}$}}}
\put(87,69){\scalebox{0.8}{$|\mathcal R|$}}

\put(-4,92){\scalebox{0.7}{5}}
\put(-4,77){\scalebox{0.7}{4}}
\put(-4,60){\scalebox{0.7}{3}}
\put(-4,45){\scalebox{0.7}{2}}
\put(-4,29){\scalebox{0.7}{1}}

\put(100,15){\scalebox{0.7}{$0$}}
\put(100,39){\scalebox{0.7}{$0.9$}}
\put(100,63){\scalebox{0.7}{$1.8$}}
\end{overpic}

\vspace{-0.48cm}

\begin{overpic}[width=0.9\linewidth]{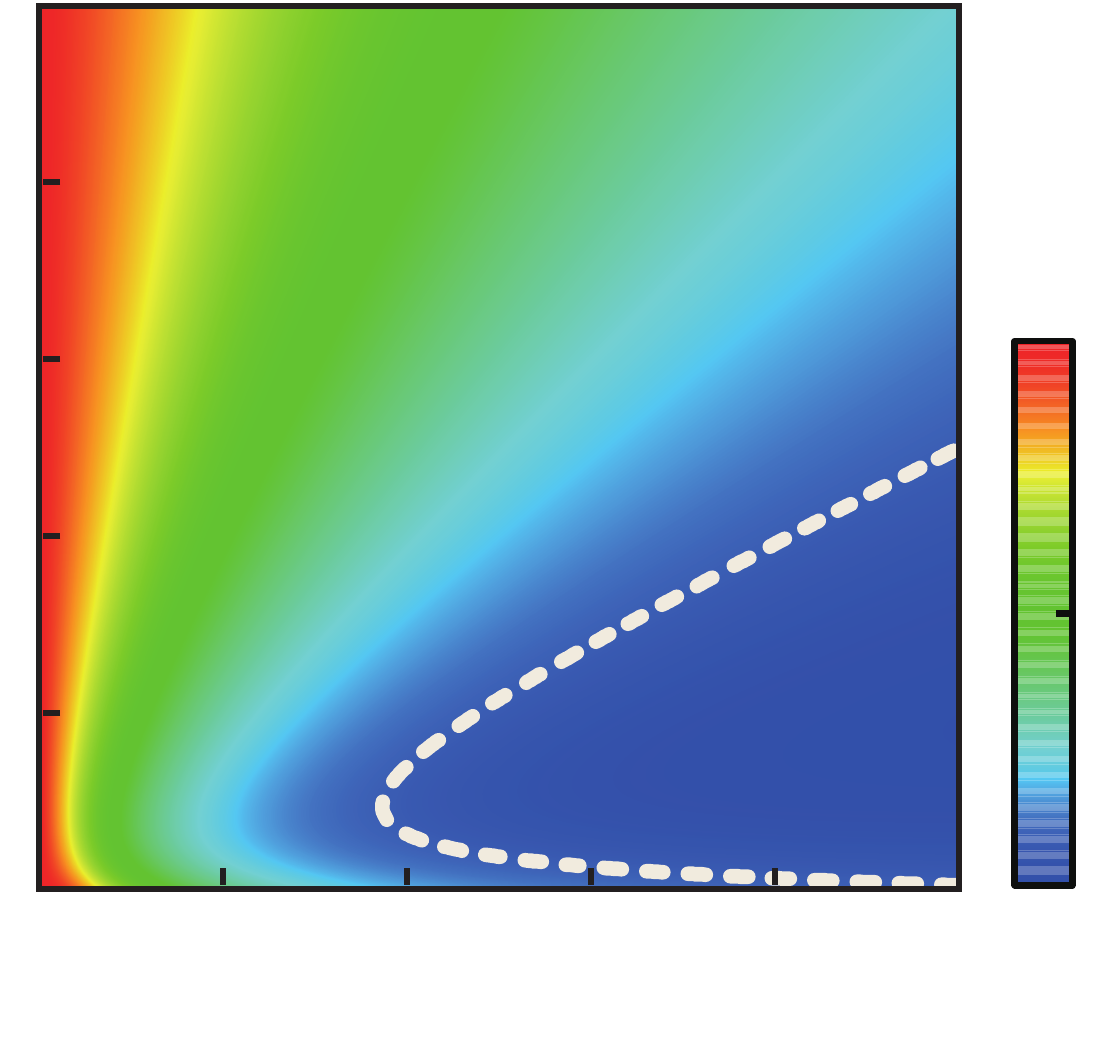}
\put(70,85){\scalebox{0.75}{\textbf{(b)}}}
\put(-15,50){\rotatebox{90}{\scalebox{0.8}{$\bm{\Delta}$}}}
\put(45,-3){\scalebox{0.8}{$\bm{\Omega}$}}
\put(91.2,69){\scalebox{0.9}{$\kappa_0$}}
\put(-4,77){\scalebox{0.7}{4}}
\put(-4,61){\scalebox{0.7}{3}}
\put(-4,45){\scalebox{0.7}{2}}
\put(-4,29){\scalebox{0.7}{1}}
\put(-3,13){\scalebox{0.7}{0}}

\put(2,7){\scalebox{0.7}{0}}
\put(18,7){\scalebox{0.7}{$1$}}
\put(35,7){\scalebox{0.7}{$2$}}
\put(52,7){\scalebox{0.7}{$3$}}
\put(68,7){\scalebox{0.7}{$4$}}
\put(84,7){\scalebox{0.7}{$5$}}

\put(100,15){\scalebox{0.7}{$0.9$}}
\put(100,38){\scalebox{0.7}{$1.6$}}
\put(100,61){\scalebox{0.7}{$2.2$}}
\end{overpic}

\end{minipage}
\hspace{-0.04\columnwidth}
\begin{minipage}[t]{0.55\columnwidth}
\vspace{0pt}
\centering
\begin{overpic}[width=1.1\linewidth]{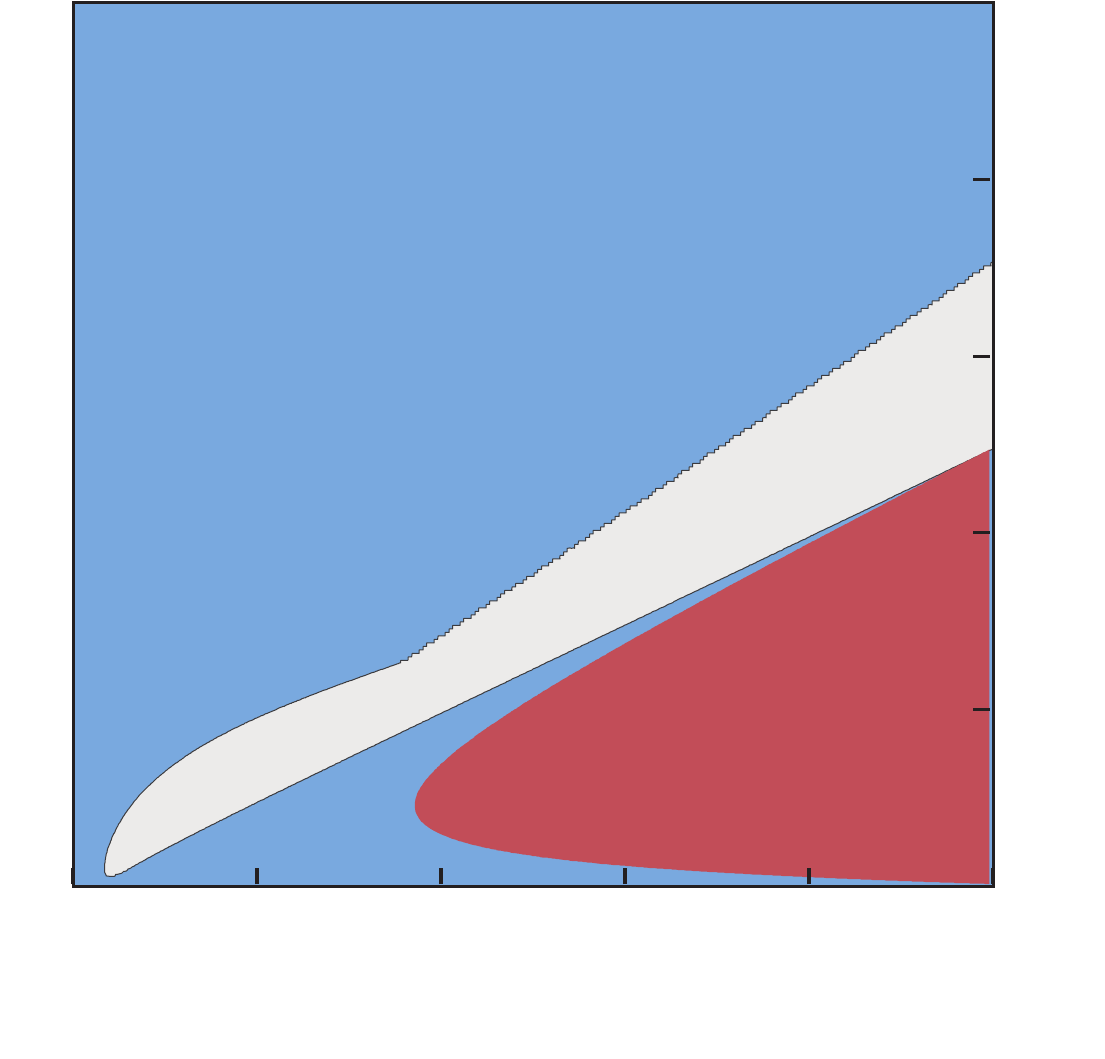}
\put(80,89){\scalebox{0.85}{\textbf{(c)}}}

\put(95,50){\rotatebox{90}{\scalebox{0.9}{$\bm{\Delta}$}}}
\put(49,6){\scalebox{0.9}{$\bm{\Omega}$}}

\put(29,69){\scalebox{0.8}{\shortstack{\textbf{Mpemba}\\ ($|\mathcal R|<1,\ \kappa_0>1$)}}}
\put(51,23){\scalebox{0.8}{\shortstack{\textbf{acceleration}\\ ($|\mathcal R|<1,\ \kappa_0\le1$)}}}
\put(50,40){\rotatebox{31}{\scalebox{0.8}{\shortstack{\\ ($|\mathcal R|\ge1,\ \kappa_0>1$)}}}}

\put(92,94){\scalebox{0.9}{5}}
\put(92,78.2){\scalebox{0.9}{4}}
\put(92,62.2){\scalebox{0.9}{3}}
\put(92,46.2){\scalebox{0.9}{2}}
\put(92,30.2){\scalebox{0.9}{1}}
\put(92,15){\scalebox{0.9}{0}}

\put(6,10){\scalebox{0.90}{0}}
\put(22,10){\scalebox{0.90}{$1$}}
\put(39.2,10){\scalebox{0.90}{$2$}}
\put(55.8,10){\scalebox{0.90}{$3$}}
\put(72.3,10){\scalebox{0.90}{$4$}}
\put(88.4,10){\scalebox{0.90}{$5$}}
\end{overpic}
\end{minipage}
}
\caption{
Parameter-induced acceleration of relaxation.
(a) Slow-mode amplitude ratio \(|\mathcal R|\). (b) Initial distance ratio
\(\kappa_0\). (c) Classification by the two diagnostics. The blue region
$|\mathcal R|<1$, $\kappa_0>1$ identifies Mpemba candidates. Parameters:
\(|\mathbf r_0|=1\), \(\theta=\arccos(-0.6)\), \(\phi=0\), \(T=0.1\),
\(\gamma=1.0\), \(N=100\). Contours use the modulus condition
\(|a+1|<|a+X|\), valid for real and complex \(\lambda_2\).}
\label{a2}
\end{figure}

Figure~\ref{a2} maps the two preparation criteria in the
$(\Omega,\Delta)$ plane. Slow-mode suppression, $|\mathcal R|<1$,
occurs over a broad region and is particularly strong for weak driving
and large detuning [Fig.~\ref{a2}(a)]. The prepared state starts farther
from stationarity, $\kappa_0>1$, over much of the plane, except in the
strong-driving, small-detuning region [Fig.~\ref{a2}(b)].
These criteria are combined in Fig.~\ref{a2}(c): the blue region
satisfies both and identifies quantum Mpemba candidates; the red region
exhibits amplitude suppression without an increased initial distance,
whereas the light-gray region exhibits an increased initial distance
without slow-mode suppression. A similar analysis for a mixed initial
state is presented in Appendix~\ref{app:accel_condition}, with the
same qualitative structure of the candidate regions.

\begin{figure*}[t]
\centering
\hspace*{0.5cm}
\begin{overpic}[width=5.5cm]{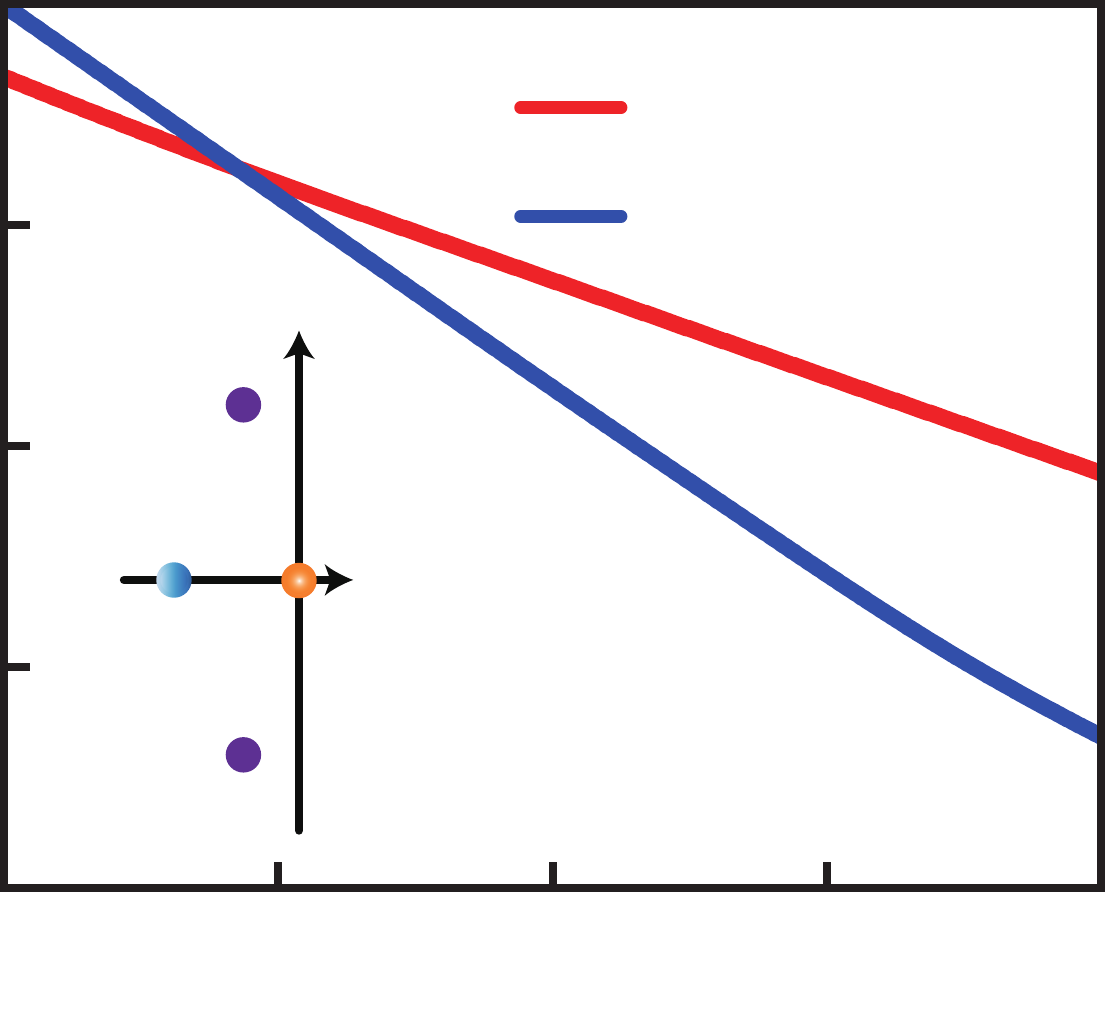}
\put(86,96){\scalebox{1.2}{(a)}}
\put(44,0){\scalebox{1.2}{$\bm{t}$}}
\put(58,82){\scalebox{1}{$D_T(\rho_t,\rho_{\text{ss}}),\rho_0$}}
\put(57,72){\scalebox{1}{$D_T(\tilde{\rho}_t,\rho_{\text{ss}}),\tilde{\rho}_N$}}
\put(-1.5,6){\scalebox{1.1}{0}}
\put(24,6){\scalebox{1.1}{2}}
\put(48,6){\scalebox{1.1}{4}}
\put(73.5,6){\scalebox{1.1}{6}}
\put(98,6){\scalebox{1.1}{8}}
\put(-14,12){\scalebox{1.1}{$10^{-4}$}}
\put(-14,30){\scalebox{1.1}{$10^{-3}$}}
\put(-14,50){\scalebox{1.1}{$10^{-2}$}}
\put(-14,70){\scalebox{1.1}{$10^{-1}$}}
\put(-10,90){\scalebox{1.1}{$10^0$}}
\put(29,58){\scalebox{1}{$\text{Im}(\lambda_i)$}}
\put(29,36){\scalebox{1}{$\text{Re}(\lambda_i)$}}
\put(21,43){\scalebox{1}{$\lambda_1$}}
\put(18,60){\scalebox{1}{$\lambda_2^*$}}
\put(18,21){\scalebox{1}{$\lambda_2$}}
\put(13,36){\scalebox{1}{$\lambda_4$}}
\end{overpic}
\vspace*{0.5cm}
\hspace{1.5cm}
\begin{overpic}[width=5.5cm]{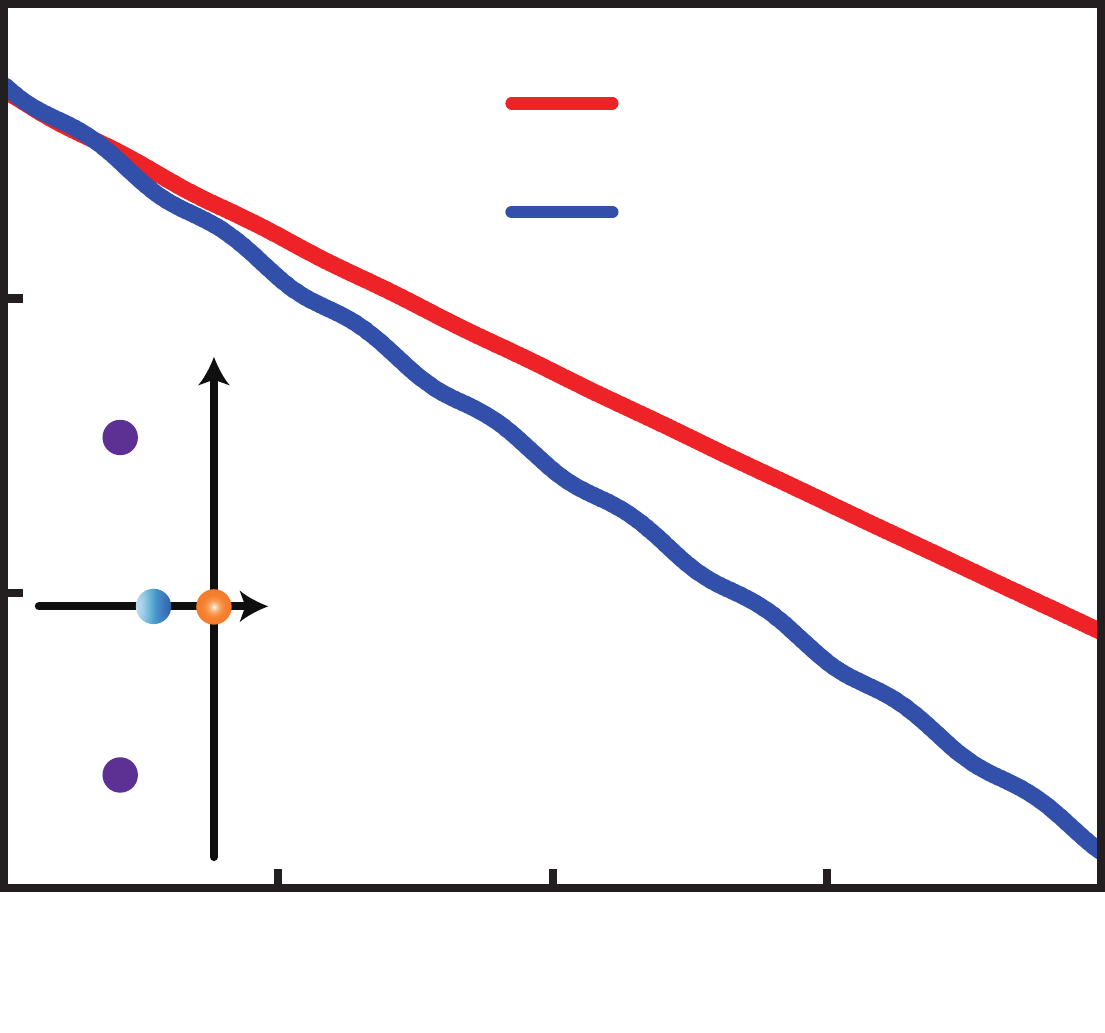}
\put(86,96){\scalebox{1.2}{(b)}}
\put(44,0){\scalebox{1.2}{$\bm{t}$}}
\put(57,82){\scalebox{1}{$D_T(\rho_t,\rho_{\text{ss}}),\rho_0$}}
\put(57,72){\scalebox{1}{$D_T(\tilde{\rho}_t,\rho_{\text{ss}}),\tilde{\rho}_N$}}
\put(-1.5,6){\scalebox{1.1}{0}}
\put(24,6){\scalebox{1.1}{2}}
\put(48,6){\scalebox{1.1}{4}}
\put(73.5,6){\scalebox{1.1}{6}}
\put(98,6){\scalebox{1.1}{8}}
\put(-14,12){\scalebox{1.1}{$10^{-3}$}}
\put(-14,39){\scalebox{1.1}{$10^{-2}$}}
\put(-14,65){\scalebox{1.1}{$10^{-1}$}}
\put(-10,90){\scalebox{1.1}{$10^0$}}
\put(21,56){\scalebox{1}{$\text{Im}(\lambda_i)$}}
\put(21,33){\scalebox{1}{$\text{Re}(\lambda_i)$}}
\put(13,40){\scalebox{1}{$\lambda_1$}}
\put(10,33){\scalebox{1}{$\lambda_2$}}
\put(9,57){\scalebox{1}{$\lambda_3^*$}}
\put(9,18){\scalebox{1}{$\lambda_3$}}
\end{overpic}
\hspace*{0.5cm}
\begin{overpic}[width=5.5cm]{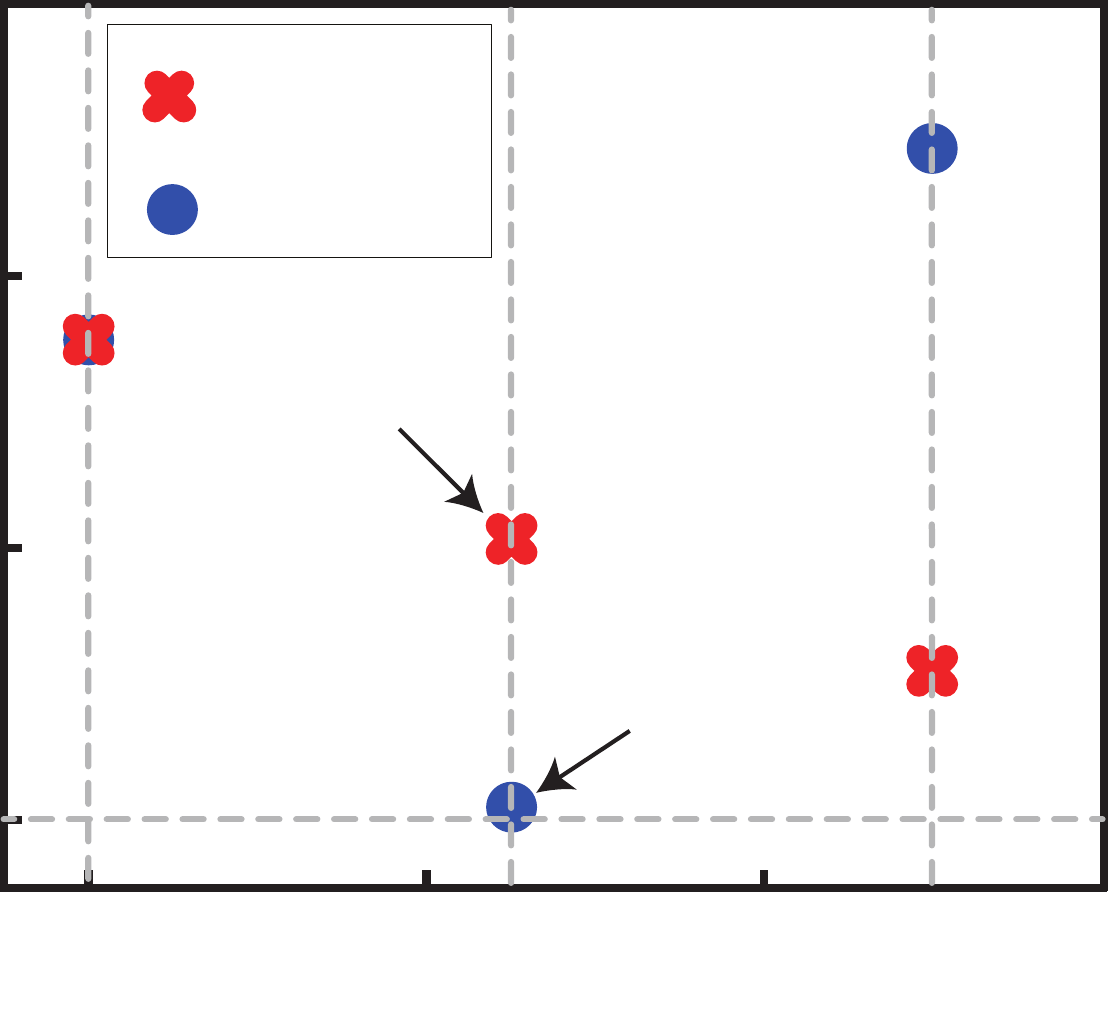}
\put(86,96){\scalebox{1.2}{(c)}}
\put(19,83){\scalebox{0.9}{$|\operatorname{Tr}(l_\nu\rho_0)|$}}
\put(19,73){\scalebox{0.9}{$|\operatorname{Tr}(l_\nu\tilde{\rho}_N)|$}}
\put(-10,90){\scalebox{1.1}{1.2}}
\put(-10,67){\scalebox{1.1}{0.8}}
\put(-10,42){\scalebox{1.1}{0.4}}
\put(-5,17){\scalebox{1.1}{0}}
\put(6,6){\scalebox{1.1}{0}}
\put(34,6){\scalebox{1.1}{0.4}}
\put(65,6){\scalebox{1.1}{0.8}}
\put(93,6){\scalebox{1.1}{1.2}}
\put(40,-2){\scalebox{1.1}{$|\mathrm{Re}(\lambda_\nu)|$}}
\put(16,58){\scalebox{1}{$|\operatorname{Tr}(l_2\rho_0)|$}}
\put(54,31){\scalebox{1}{$|\operatorname{Tr}(l_2\tilde{\rho}_N)|$}}
\end{overpic}
\hspace{1.5cm}
\begin{overpic}[width=5.5cm]{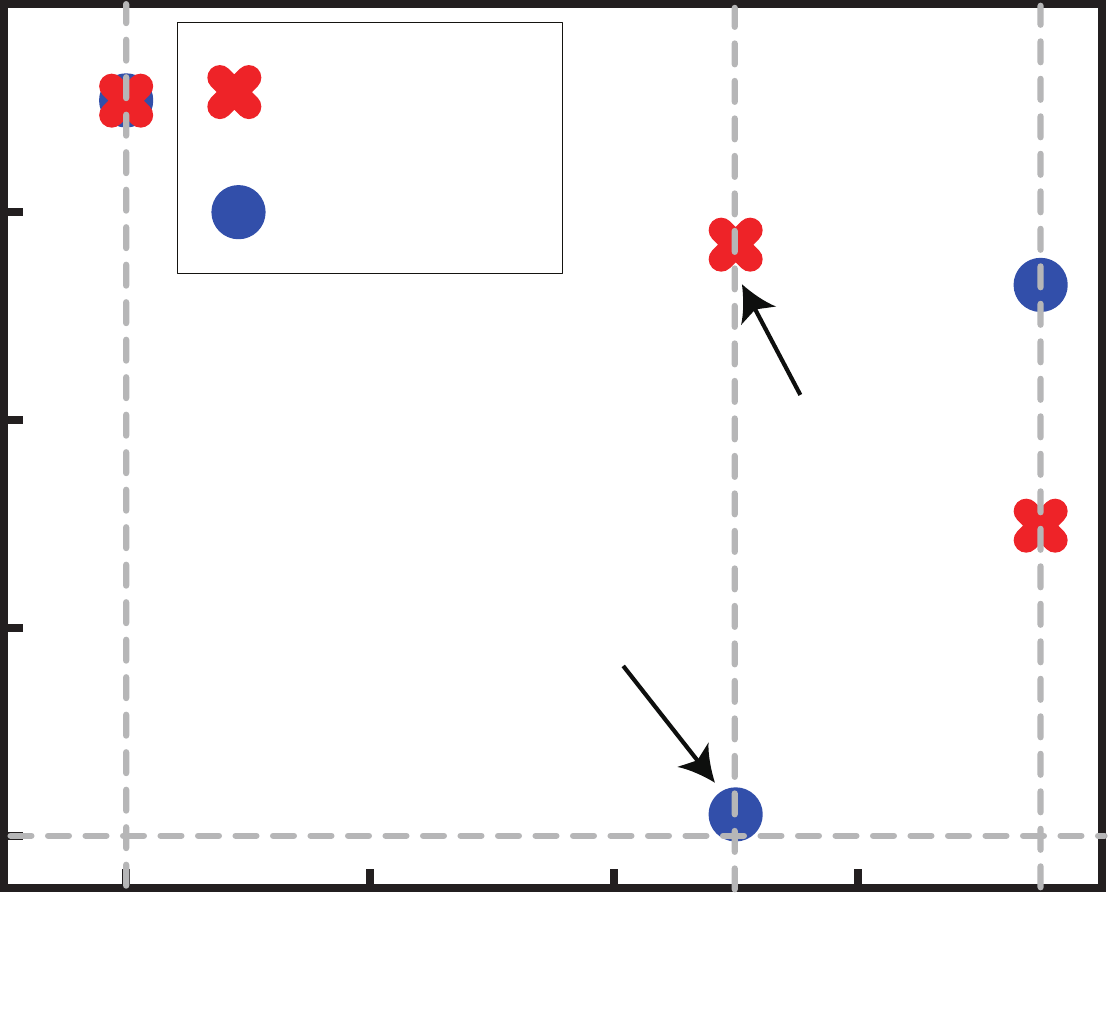}
\put(86,96){\scalebox{1.2}{(d)}}
\put(25,83){\scalebox{0.9}{$|\operatorname{Tr}(l_\nu\rho_0)|$}}
\put(25,72){\scalebox{0.9}{$|\operatorname{Tr}(l_\nu\tilde{\rho}_N)|$}}
\put(9,6){\scalebox{1.1}{0}}
\put(28,6){\scalebox{1.1}{0.2}}
\put(50,6){\scalebox{1.1}{0.4}}
\put(72,6){\scalebox{1.1}{0.6}}
\put(92,6){\scalebox{1.1}{0.8}}
\put(-5,16){\scalebox{1.1}{0}}
\put(-10,33){\scalebox{1.1}{0.2}}
\put(-10,53){\scalebox{1.1}{0.4}}
\put(-10,71){\scalebox{1.1}{0.6}}
\put(-10,90){\scalebox{1.1}{0.8}}
\put(40,-2){\scalebox{1.2}{$|\mathrm{Re}(\lambda_\nu)|$}}
\put(68,52){\scalebox{1}{$|\operatorname{Tr}(l_2\rho_0)|$}}
\put(40,35){\scalebox{1}{$|\operatorname{Tr}(l_2\tilde{\rho}_N)|$}}
\end{overpic}

\caption{Quantum Mpemba effect in a driven dissipative qubit with complex
(a), (c) and real (b), (d) spectral gaps. (a), (b) Trace distance to
\(\rho_{\mathrm{ss}}\) for \(\rho_0\) (red) and \(\tilde\rho_N\) (blue);
insets: spectrum of \(\mathcal L\). (c), (d) Modal weights
\(|\mathrm{Tr}(l_\nu\rho)|\) vs \(|\mathrm{Re}(\lambda_\nu)|\). The
preparation suppresses the slowest modal weights, including both members of
the complex-conjugate pair in (c). The parameters are
\((\Omega,\Delta)=(0.1,2)\) for panels (a) and (c), and
\((3.2,0.05)\) for panels (b) and (d). Other parameters as in Fig.~\ref{a2}.
Time starts after preparation.}
\label{DT}
\end{figure*}

Figures~\ref{DT}(a) and~\ref{DT}(b) compare relaxation from the original state $\rho_0$ and from the prepared state $\tilde{\rho}_N$. In Fig.~\ref{DT}(a), the long-time dynamics of the original state are governed by a slow complex-conjugate pair,
\(
\rho_t \simeq \rho_{\mathrm{ss}}
+ e^{\mathrm{Re}(\lambda_2)t}
\Big[
\mathrm{Tr}(l_2 \rho_0)\, r_2 e^{ i\,\mathrm{Im}(\lambda_2)t }
+ \mathrm{Tr}(l_2^\dagger \rho_0)\, r_2^\dagger e^{- i\,\mathrm{Im}(\lambda_2)t }
\Big].
\)
The crossing of the curves demonstrates the quantum Mpemba effect: the prepared state starts farther from equilibrium but reaches the steady state sooner because the slow modes are suppressed.
The same conclusion holds when the slowest eigenvalue is real, as shown in Fig.~\ref{DT}(b), where
\(
\rho_t \simeq \rho_{\mathrm{ss}} + \mathrm{Tr}(l_2\rho_0)\, r_2 e^{\lambda_2 t}.
\)

{The mechanism is displayed directly by the modal weights
$|\mathrm{Tr}(l_\nu\rho)|$, plotted against the decay rates
$|\mathrm{Re}(\lambda_\nu)|$ [Figs.~\ref{DT}(c) and~\ref{DT}(d)].} After preparation, the slowest-mode weights are strongly reduced, while faster modes retain appreciable weight. In Fig.~\ref{DT}(c), for example, $|\mathrm{Tr}(l_2\tilde{\rho}_N)| \approx 0$ and $|\mathrm{Tr}(l_2^\dagger\tilde{\rho}_N)| \approx 0$, so faster channels can dominate the intermediate relaxation. A nonzero
residual slow weight retains the original ultimate decay exponent.
The same argument applies to the real-eigenvalue case in Fig.~\ref{DT}(d).

For the driven qubit, the limit \(\Omega=0\) provides a transparent
Davies-type reference: the energy and bare bases coincide, and the
Liouvillian separates into population $l_{\rm pop}$ and coherence sectors $l_{\rm coh}$. The slowest modes are coherences
decaying at rate \(\gamma/2\), while populations decay at rate
\(\gamma\). A nonselective measurement preparation diagonal in this basis
therefore satisfies \([\tilde\rho_N,H]=0\), even for mixed states, and is
orthogonal to the slow coherence modes:
${\rm Tr}(l_{\rm coh}\tilde\rho_N)=0$.
The relaxation then bypasses the slow coherence tail and proceeds through
the faster population channel. For the finite-$N$ geodesic sequence,
exact diagonality is reached only in the limit or after an additional
final dephasing; otherwise a residual slow coherence remains.

\subsection{Coupled multiqubit system}
\label{sec:multiqubit}

\subsubsection{Measurement-based state preparation}
\label{subsubsec:multiqubit_preparation}

While the single-qubit model establishes the steering mechanism in the
minimal setting, the multiqubit problem probes symmetry-resolved
coherence sectors of the Liouvillian that have no direct counterpart in
single-system state steering~\cite{Pec06}. We consider \(M\) coupled
qubits with Hilbert-space dimension \(d=2^M\), governed by the
transverse-field XY Hamiltonian~\cite{Rib07,Dus04,Kam07}
\begin{equation}
H =
J \sum_{i<j}
\left(
\sigma_x^{(i)}\sigma_x^{(j)}
+
\sigma_y^{(i)}\sigma_y^{(j)}
\right)
+
\frac{\Omega}{2}\sum_{i=1}^{M}\sigma_z^{(i)}
+
\frac{\Omega_d}{2}\sum_{i=1}^{M}\sigma_x^{(i)} .
\label{eq:many_qubit_H}
\end{equation}
Here $\{|0\rangle,|1\rangle\}$ denotes the computational basis of each
qubit, with $\sigma_z|1\rangle=|1\rangle$ and
$\sigma_z|0\rangle=-|0\rangle$, so that $|0\rangle$ is the local
ground state of the decay channel introduced below; the Pauli
operators $\sigma_{x,y,z}^{(i)}$ act on the $i$th qubit, and the
formalism applies to any effective two-level system. As the target we take the
fully excited product state
$|\tau\rangle=|e\rangle\equiv|1\rangle^{\otimes M}=|11\cdots 1\rangle$.
This target is far from the steady state, which lies close to
\(|0\rangle^{\otimes M}\) for zero-temperature dissipation, favoring
\(\kappa_0>1\); for a pure input, the prepared state approaches this diagonal
target as $N$ increases. For mixed inputs, residual slow-mode
overlaps must be evaluated for the full prepared state.

{We use the
same \(N\)-measurement, \(N+1\)-unitary protocol with
\(U_k=\exp(-iH_{\rm eff}\Delta t_k)\).
The global measurement bases are constructed by rotating only the
two-dimensional subspace that connects the dominant component of the
initial state to the effective target direction. Let \(|a\rangle\) denote the dominant eigenstate of
the initial density matrix \(\rho_0\), namely the eigenstate carrying the
largest population. We choose another normalized state \(|b\rangle\), with
\(\langle a|b\rangle=0\), such that the effective target direction \(|e_{\rm {eff}}\rangle=U_{\rm tot}^{\dagger}|e\rangle\) lies in
the two-dimensional steering subspace
\begin{equation}
\mathcal H_2
=
{\rm span}\{|a\rangle,|b\rangle\}.
\end{equation}
Equivalently, the effective target direction can be reached by a rotation within
\(\mathcal H_2\).

Within this restricted construction, the Fubini--Study geodesic connecting two pure states lies
entirely within their two-dimensional span, so the shortest
path from \(|a\rangle\) to \(|e_{\rm eff}\rangle\) is contained in
\(\mathcal H_2\), and the equal-step construction of Sec.~\ref{subsec:example_qubit} applies
to the Bloch direction of the state's component in this
effective two-level subspace. This does not establish global optimality
for a general mixed multilevel input. Here
$\delta\theta_{\rm eff}=2\arccos|\langle a|e_{\rm eff}\rangle|$
is the effective Bloch angle; phases are chosen so that
$|e_{\rm eff}\rangle=\cos(\delta\theta_{\rm eff}/2)|a\rangle
+\sin(\delta\theta_{\rm eff}/2)|b\rangle$. 
The following measurement bases are defined in the rotating
frame. The laboratory projectors are
$\tilde\Pi_{\mathbf a_k}^{(\mu)}
=V_k|\Phi_\mu^{(k)}\rangle\langle\Phi_\mu^{(k)}|V_k^\dagger$.
At the \(k\)th measurement step, the two basis vectors spanning
\(\mathcal H_2\) are rotated according to
\begin{equation}
\theta_k
=
\frac{k}{2(N+1)}
\delta\theta_{\rm eff},
\qquad
k=1,\ldots,N ,
\label{eq:many_qubit_theta}
\end{equation}
with
\begin{align}
|\Phi_1^{(k)}\rangle
&=
\cos\theta_k\,|a\rangle
+
\sin\theta_k\,|b\rangle ,
\label{eq:many_qubit_phi1}
\\
|\Phi_2^{(k)}\rangle
&=
-\sin\theta_k\,|a\rangle
+
\cos\theta_k\,|b\rangle .
\label{eq:many_qubit_phi2}
\end{align}
The remaining \(d-2\) basis vectors are kept fixed in the orthogonal
complement of \(\mathcal H_2\),
\begin{equation}
|\Phi_j^{(k)}\rangle
=
|\chi_j\rangle,
\qquad
j=3,\ldots,d,
\label{eq:many_qubit_fixed}
\end{equation}
where
\(\{|\chi_3\rangle,\ldots,|\chi_d\rangle\}\)
form an orthonormal basis of
\(\mathcal H_2^\perp\), obtained by projecting the computational basis vectors, in binary
order, onto $\mathcal H_2^\perp$ and applying Gram--Schmidt
orthogonalization, discarding zero residual vectors.  The complete measurement basis is
therefore
\begin{equation}
\mathcal B_k
=
\{
|\Phi_1^{(k)}\rangle,
|\Phi_2^{(k)}\rangle,
|\Phi_3^{(k)}\rangle,
\ldots,
|\Phi_d^{(k)}\rangle
\},
\label{eq:many_qubit_basis}
\end{equation}
and the corresponding nonselective measurement channel is
\begin{equation}
\mathcal M_{\mathbf a_k}[\rho]
=
\sum_{\mu=1}^{d}
\Pi_{\mathbf a_k}^{(\mu)}
\rho
\Pi_{\mathbf a_k}^{(\mu)},
\qquad
\Pi_{\mathbf a_k}^{(\mu)}
=
|\Phi_\mu^{(k)}\rangle
\langle\Phi_\mu^{(k)}|.
\label{eq:many_qubit_measurement}
\end{equation}
They satisfy
$\sum_{\mu=1}^{d}\Pi_{\mathbf a_k}^{(\mu)}
=
I_d $.
The construction has the same geometric interpretation as in the
single-qubit case. The measurement basis advances in equal angular
increments along the geodesic from the dominant initial direction to
the effective target, while the complement basis is fixed. Its initial coherences are
erased by the first complete measurement. For a mixed input, the
final unitary can leave coherence in the computational basis, so
this construction does not guarantee a globally diagonal output.

\textit{Remark on experimental implementation.}---Each
global nonselective measurement in the basis $\mathcal B_k$ is
operationally a mid-circuit readout of all qubits in the computational
basis, with the outcomes discarded, conjugated by the basis-change
unitary $W_k$ that maps the computational basis onto $\mathcal B_k$:
$\mathcal M_{\mathbf a_k}[\rho]=W_k\,\Lambda_{\rm com}[W_k^\dagger\rho
W_k]\,W_k^\dagger$, where $\Lambda_{\rm com}$ denotes complete dephasing
in the computational basis. Writing $W_k=G(\theta_k)W_0$, where $W_0$
maps two computational basis states onto $\{|a\rangle,|b\rangle\}$ and
$G(\theta_k)$ is the two-level (Givens) rotation
[Eqs.~\eqref{eq:many_qubit_phi1}--\eqref{eq:many_qubit_phi2}] acting inside
$\mathcal H_2$, consecutive rotating-frame bases differ by the
fixed-angle rotation $G(\delta\theta_{\rm eff}/[2(N+1)])$.
The physical basis-change unitary is $V_kW_k$ and
also includes the known cumulative coherent evolution.

{Implementing this construction requires knowledge of $|a\rangle$, the
effective target, and the chosen measurement bases. The basis changes
are generally global in the physical-qubit representation. A two-level
rotation in $\mathcal H_2$ can therefore require local and entangling
gates, with a cost that depends on $M$ and the input state. The fixed
geometric increment does not imply size-independent circuit depth or
negligible preparation overhead.}}

\subsubsection{Slow-mode suppression and relaxation dynamics}
\label{subsubsec:multiqubit_relaxation}

We use a local-dissipator model: with the Hamiltonian in Eq.~\eqref{eq:many_qubit_H}, the local decay operators are \(L_j=\sqrt{\gamma_j}\,\sigma_-^{(j)}\).
Since these jumps are defined in the bare-qubit basis rather than through
Bohr-frequency-resolved transitions of the interacting Hamiltonian, the
Liouvillian is not a Davies generator in the global energy basis. The
transverse field \(\Omega_d\) breaks residual symmetries and makes
population--coherence mixing of the slow mode visible. The $\Omega_d=0$ case provides an excitation-conserving local-decay
reference.

\begin{figure*}[t]
\centering
\hspace*{0.5cm}
\begin{overpic}[height=5.5cm]{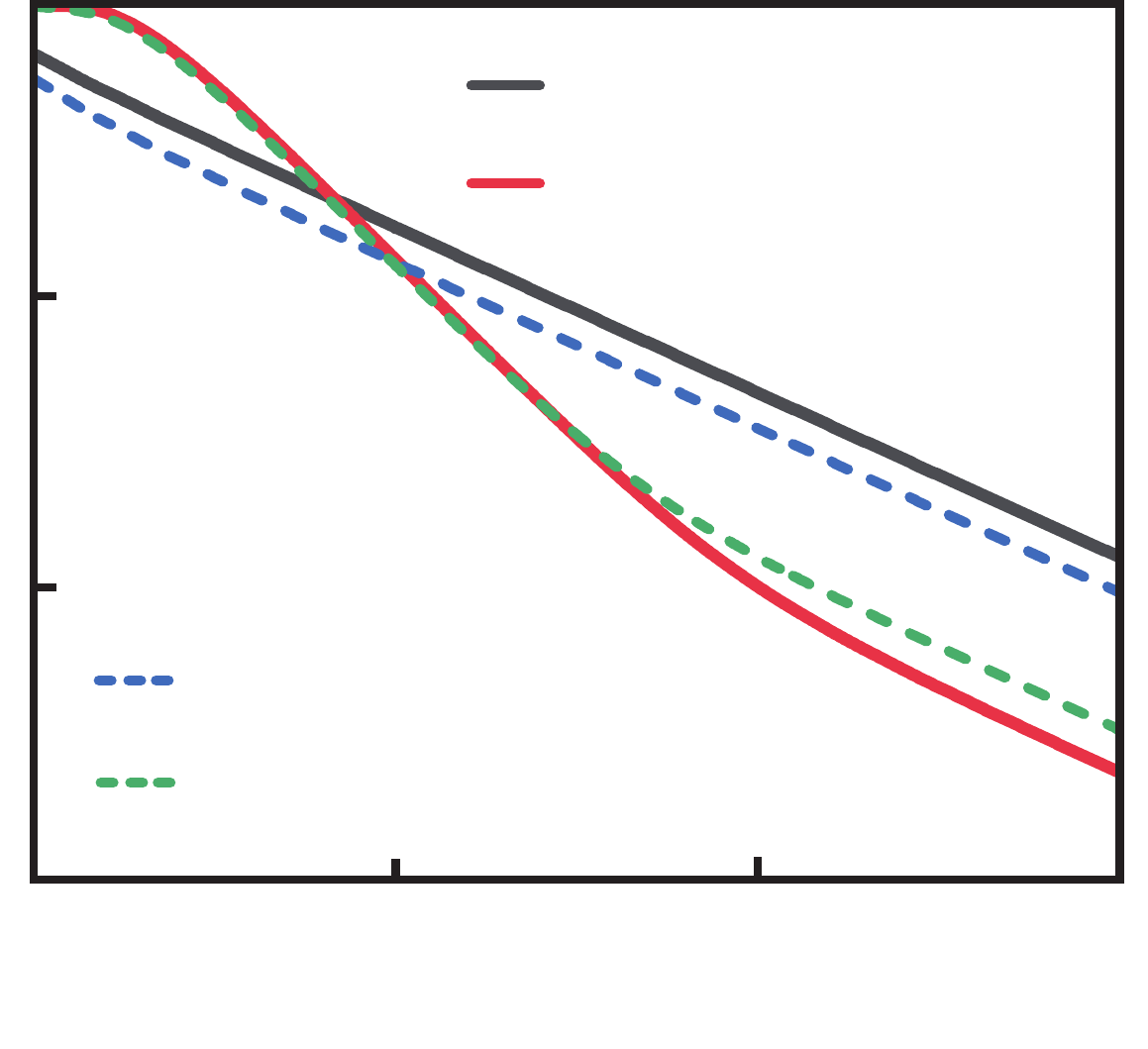}
\put(86,95){\scalebox{1.2}{(a)}}
\put(44,2){\scalebox{1.2}{$\bm{t}$}}
\put(50,83){\scalebox{1}{$D_T(\rho_t,\rho_{\text{ss}}),\rho_0$}}
\put(50,74){\scalebox{1}{$D_T(\tilde{\rho}_t,\rho_{\text{ss}}),\tilde{\rho}_N$}}
\put(17,31){\scalebox{1}{$D_T(\rho_t',\rho_{\text{ss}}),\rho_0'$}}
\put(17,21.5){\scalebox{1}{$D_T(\tilde{\rho}_t',\rho_{\text{ss}}),\tilde{\rho}_N'$}}
\put(1.5,8){\scalebox{1.1}{0}}
\put(33,8){\scalebox{1.1}{5}}
\put(63,8){\scalebox{1.1}{10}}
\put(93,8){\scalebox{1.1}{15}}
\put(-11,15){\scalebox{1.1}{$10^{-6}$}}
\put(-11,40){\scalebox{1.1}{$10^{-4}$}}
\put(-11,65){\scalebox{1.1}{$10^{-2}$}}
\put(-7,90){\scalebox{1.1}{$10^0$}}
\end{overpic}
\hspace{1.4cm}
\begin{overpic}[height=5.5cm]{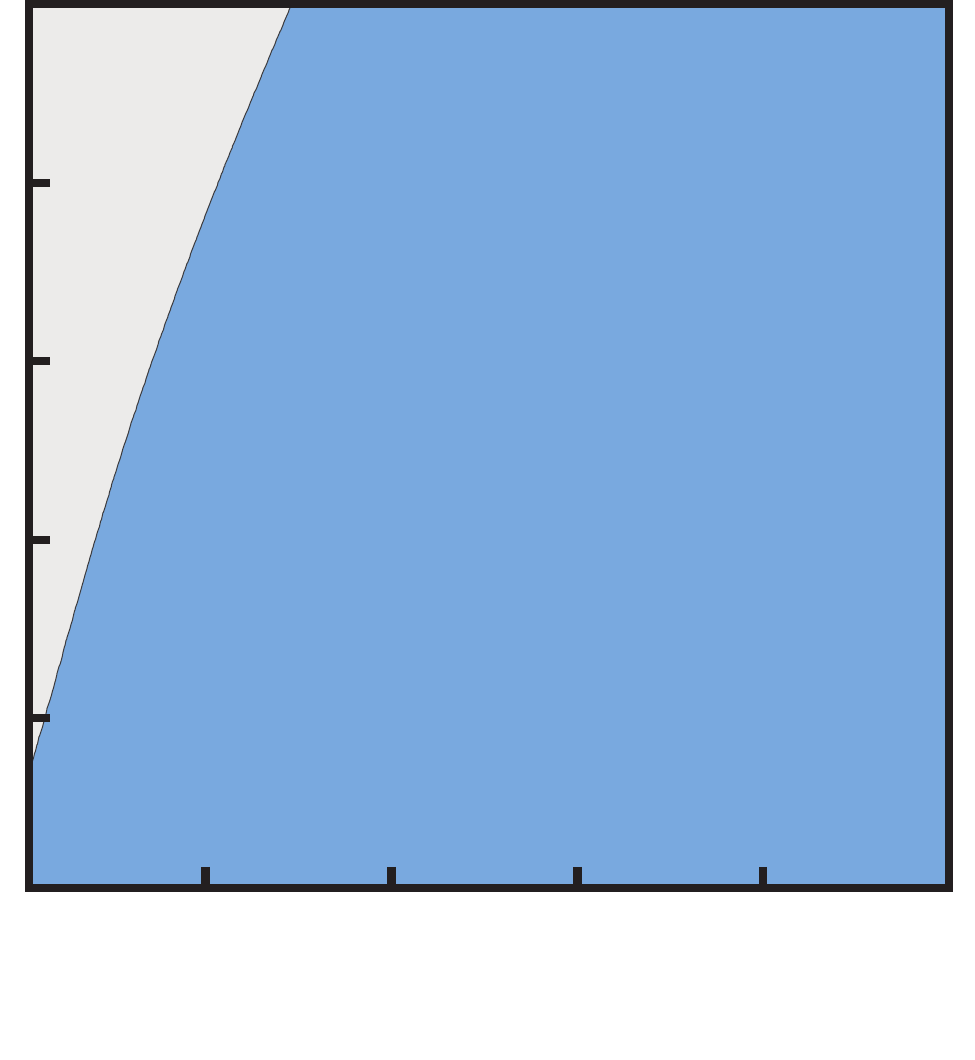}
\put(78,103){\scalebox{1.2}{(b)}}
\put(30,40){\scalebox{1}{\shortstack{\textbf{Mpemba}\\ ($|\mathcal{R}|<1,\ \kappa_0>1$)}}}
\put(-20,45){\rotatebox{90}{\scalebox{1.1}{$|\mathbf r_2(0)|$}}}
\put(45,0){\scalebox{1.1}{$|\mathbf r_1(0)|$}}
\put(-6,96){\scalebox{1.1}{1}}
\put(-8.5,80){\scalebox{1.1}{0.8}}
\put(-8.5,63){\scalebox{1.1}{0.6}}
\put(-8.5,47){\scalebox{1.1}{0.4}}
\put(-8.5,29){\scalebox{1.1}{0.2}}
\put(-6,12){\scalebox{1.1}{0}}
\put(0,8){\scalebox{1.1}{0}}
\put(16,8){\scalebox{1.1}{$0.2$}}
\put(33.2,8){\scalebox{1.1}{$0.4$}}
\put(52.8,8){\scalebox{1.1}{$0.6$}}
\put(70.3,8){\scalebox{1.1}{$0.8$}}
\put(89,8){\scalebox{1.1}{$1$}}
\end{overpic}\\[0.25cm]
\hspace*{0.5cm}
\begin{overpic}[height=5.5cm]{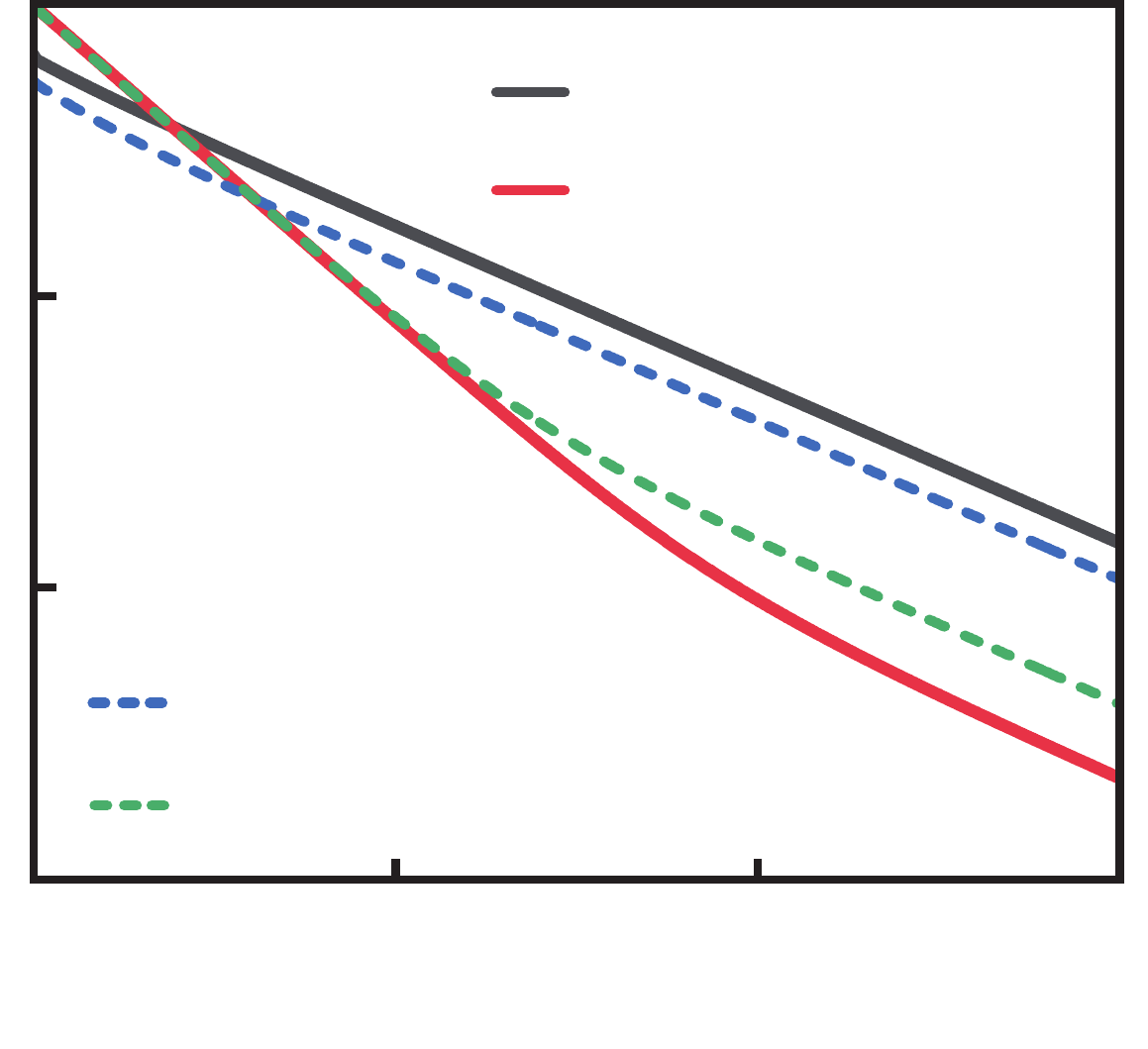}
\put(86,95){\scalebox{1.2}{(c)}}
\put(44,2){\scalebox{1.2}{$\bm{t}$}}
\put(53,82){\scalebox{1}{$D_T(\rho_t,\rho_{\text{ss}}),\rho_0$}}
\put(53,73){\scalebox{1}{$D_T(\tilde{\rho}_t,\rho_{\text{ss}}),\tilde{\rho}_N$}}
\put(17,29){\scalebox{1}{$D_T(\rho_t',\rho_{\text{ss}}),\rho_0'$}}
\put(17,20.5){\scalebox{1}{$D_T(\tilde{\rho}_t',\rho_{\text{ss}}),\tilde{\rho}_N'$}}
\put(1.5,8){\scalebox{1.1}{0}}
\put(33,8){\scalebox{1.1}{5}}
\put(63,8){\scalebox{1.1}{10}}
\put(93,8){\scalebox{1.1}{15}}
\put(-11,15){\scalebox{1.1}{$10^{-6}$}}
\put(-11,40){\scalebox{1.1}{$10^{-4}$}}
\put(-11,65){\scalebox{1.1}{$10^{-2}$}}
\put(-7,90){\scalebox{1.1}{$10^0$}}
\end{overpic}
\hspace{1.4cm}
\begin{overpic}[height=5.5cm]{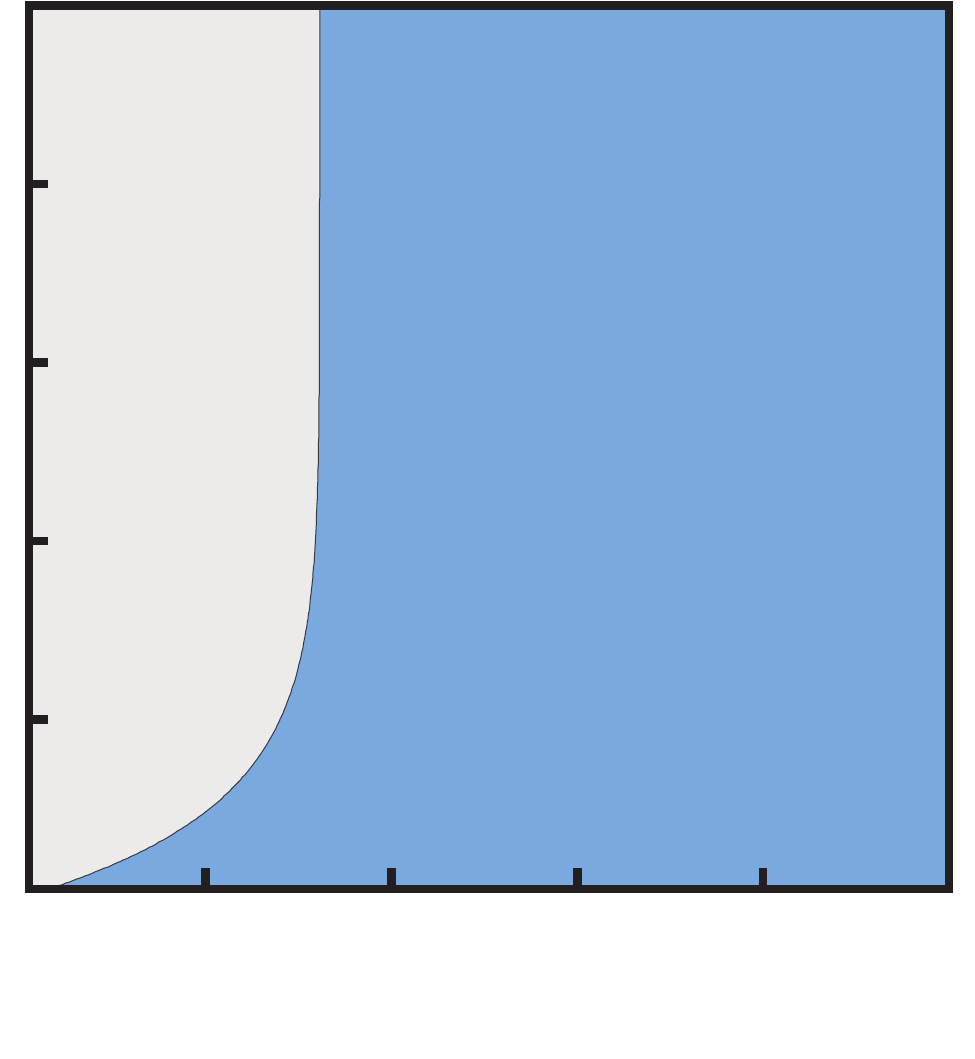}
\put(78,103){\scalebox{1.2}{(d)}}
\put(32,40){\scalebox{1}{\shortstack{\textbf{Mpemba}\\ ($|\mathcal{R}|<1,\ \kappa_0>1$)}}}
\put(-20,45){\rotatebox{90}{\scalebox{1.1}{$|\mathbf r_2(0)|$}}}
\put(45,0){\scalebox{1.1}{$|\mathbf r_1(0)|$}}
\put(-6,96){\scalebox{1.1}{1}}
\put(-8.5,80){\scalebox{1.1}{0.8}}
\put(-8.5,63){\scalebox{1.1}{0.6}}
\put(-8.5,47){\scalebox{1.1}{0.4}}
\put(-8.5,29){\scalebox{1.1}{0.2}}
\put(-6,12){\scalebox{1.1}{0}}
\put(0,8){\scalebox{1.1}{0}}
\put(16,8){\scalebox{1.1}{$0.2$}}
\put(33.2,8){\scalebox{1.1}{$0.4$}}
\put(52.8,8){\scalebox{1.1}{$0.6$}}
\put(70.3,8){\scalebox{1.1}{$0.8$}}
\put(89,8){\scalebox{1.1}{$1$}}
\end{overpic}

\caption{
Trace-distance dynamics and Mpemba regions for \(M=6\): {driven local decay}
(a,b) and {zero-temperature excitation-conserving benchmark} (c,d). In (a,c), solid and
dashed curves correspond to two product initial states, differing only in
the first qubit, and to their prepared states. In (b,d), blue denotes
\(|\mathcal R|<1\) and \(\kappa_0>1\). Parameters:
\((|\mathbf r_1(0)|,\theta_1,\phi_1)=(1,\arccos(-0.6),0)\) and
\((0.75,\arccos(-0.8),0)\), \(\Omega=1.2\), \(J=0.5\), \(N=100\), and \(T=0.1\); {driven}: \((\gamma_1=1,\gamma_{j>1}=1.2,\Omega_d=0.1)\); {excitation-conserving benchmark}:
\((\gamma_j=1~\forall j,\,\Omega_d=0)\).
}
\label{fig:mpemba_regions}
\end{figure*}

The results for \(M=6\) qubits are summarized in
Fig.~\ref{fig:mpemba_regions}. Two product initial states that differ
only in the first qubit are compared in Fig.~\ref{fig:mpemba_regions}(a),
\[
\rho_0
=
\rho_s^{(1)}
\otimes
(|0\rangle\langle0|)^{\otimes5},
\qquad
\rho_0'
=
\rho_s^{(1)\prime}
\otimes
(|0\rangle\langle0|)^{\otimes5}.
\]
We take
\((|\mathbf r_1(0)|,\theta_1,\phi_1)
=
(1,\arccos(-0.6),0)\)
for \(\rho_0\), and
\((0.75,\arccos(-0.8),0)\)
for \(\rho_0'\).
The black solid and blue dashed curves show the unprepared dynamics from
\(\rho_0\) and \(\rho_0'\), respectively, whereas the red solid and green
dashed curves show the corresponding measurement-prepared dynamics.
The crossings between the prepared and unprepared trajectories signal
the Mpemba effect, demonstrating that the acceleration persists for both
pure and mixed initial states.
Figure~\ref{fig:mpemba_regions}(b) maps this robustness in the plane of
initial-state purities, with \( |\mathbf r_1(0)| \) for the first qubit and
\( |\mathbf r_2(0)| \) for the remaining five. The blue regions satisfying the plotted diagnostics cover
a large fraction of parameter space.
{The scans use fixed directions
$\mathbf n_1=(0.8,0,-0.6)$ and $\mathbf n_{j>1}=-\hat{\mathbf z}$,
with $101\times101$ cell-center samples in $0<r_1,r_2<1$.
The dominant eigenvector is chosen continuously from positive
$r_1,r_2$ at degeneracies; the fully mixed input is unchanged
and excluded from the strict criteria. The excitation-conserving
map uses the collective slow eigenmode.
The full Hilbert space is retained in these calculations.}

{
The underlying mechanism can be understood directly from the slow-mode
amplitude ratio $\mathcal R$. For fixed dissipative dynamics and fixed
\(\rho_0\), the preparation protocol modifies \(\mathcal R\) only
through the overlap
\({\rm Tr}(l_2\tilde\rho_N)\).
To expose the role of measurement-induced dephasing, we decompose both
operators in the computational basis as
\[
l_2=l_2^{\rm diag}+l_2^{\rm off},
\qquad
\tilde\rho_N
=
\tilde\rho_N^{\rm diag}
+
\tilde\rho_N^{\rm off}.
\]
The diagonal and off-diagonal sectors are orthogonal under the trace
pairing, so that
\begin{equation}
{\rm Tr}(l_2\tilde\rho_N)
=
{\rm Tr}
\left(
l_2^{\rm diag}\tilde\rho_N^{\rm diag}
\right)
+
{\rm Tr}
\left(
l_2^{\rm off}\tilde\rho_N^{\rm off}
\right).
\label{eq:R_split_main}
\end{equation}
We normalize the left eigenoperator by \(\|l_2\|_F=1\) and define
\(\epsilon_c\equiv\|l_2^{\rm diag}\|_F\), where
\(\|A\|_F=\sqrt{\mathrm{Tr}(A^\dagger A)}\)
is the Frobenius norm.
Using the Cauchy--Schwarz inequality and
$\left\|
\tilde\rho_{N}^{\rm diag}
\right\|_F
\le
\|\tilde\rho_N\|_F
\le
1$, Eq.~\eqref{eq:R_split_main} gives
\begin{equation}
\left|
{\rm Tr}(l_2\tilde\rho_N)
\right|
\le
\epsilon_c
+
\sqrt{1-\epsilon_c^2}\,
\|\tilde\rho_N^{\rm off}\|_F .
\label{eq:R_bound_main}
\end{equation}
Consequently, Eq.~\eqref{rr0} has
$|\mathcal R(N)|
\le
\frac{
\epsilon_c
+
\sqrt{1-\epsilon_c^2}
\left\|
\tilde\rho_{N}^{\rm off}
\right\|_F
}{
\left|
{\rm Tr}(l_2\rho_0)
\right|
}$.
A looser but more transparent form is
\begin{equation}
|\mathcal R(N)|
\le
\frac{
\epsilon_c
+
\left\|
\tilde\rho_{N}^{\rm off}
\right\|_F
}{
\left|
{\rm Tr}(l_2\rho_0)
\right|
}.
\label{eq:SM_comp_bound_simple}
\end{equation}
Equation~\eqref{eq:SM_comp_bound_simple} identifies two contributions
to the residual slow-mode amplitude: the diagonal weight of the slowest
left eigenoperator, quantified by \(\epsilon_c\), and the residual
off-diagonal weight of the prepared state,
\(\|\tilde\rho_N^{\rm off}\|_F\). The latter term must be evaluated after the complete preparation;
dephasing in changing bases does not by itself guarantee its
monotonic decrease in the computational basis. The quantity
$\epsilon_c$ measures the diagonal component of the selected slow mode. Strong
suppression is therefore expected when both
\(\epsilon_c\ll1\) and
\(\|\tilde\rho_N^{\rm off}\|_F\ll1\). For
\(\epsilon_c=0\), the residual overlap is determined entirely by the
off-diagonal component of the prepared state and vanishes when
\(\tilde\rho_N^{\rm off}=0\).

This limiting situation is realized by the zero-temperature
excitation-conserving benchmark shown in
Figs.~\ref{fig:mpemba_regions}(c) and
\ref{fig:mpemba_regions}(d).
For \(\Omega_d=0\), the Hamiltonian commutes with the total excitation
number
\begin{equation}
\hat n
=
\sum_{i=1}^{M}
\frac{1+\sigma_z^{(i)}}{2},
\qquad
[\hat n,H]=0,
\label{eq:n_main}
\end{equation}
and the Liouvillian possesses a weak \(U(1)\) symmetry.
The operator space then decomposes into sectors labeled by
$\Delta n
=
n(\boldsymbol s)-n(\boldsymbol s')$,
where
\(|\boldsymbol s\rangle\langle\boldsymbol s'|\)
is a computational-basis matrix unit and
\(n(\boldsymbol s)=\langle\boldsymbol s|\hat n|\boldsymbol s\rangle
=\sum_{i=1}^{M}(1+\langle s_i|\sigma_z^{(i)}|s_i\rangle)/2
=\sum_{i=1}^{M}s_i\)
is the number of excitations in \(\boldsymbol s\).
The \(\Delta n=0\) sector contains populations and
same-excitation coherences, whereas
\(\Delta n=\pm1\) contains the single-quantum coherences.

For uniform decay rates \(\gamma_j=\gamma\), the slowest nonstationary modes in the $\Delta n=0$ sector
relax at rate $\gamma$, while the slowest Liouvillian modes lie
in the \(\Delta n=\pm1\) coherence sectors and decay at rate
\(\gamma/2\). Consequently, the corresponding left eigenoperator has
no diagonal component,
$l_2^{\rm diag}=0$,
and hence
$\epsilon_c=0$.
A state prepared to be diagonal in the computational basis is therefore
orthogonal to the slowest coherence sector,
\begin{equation}
{\rm Tr}(l_2\tilde\rho_N)
=
{\rm Tr}
\left(
l_2^{\rm off}\tilde\rho_N^{\rm off}
\right)
\longrightarrow 0,
\qquad
|\mathcal R|\longrightarrow0 .
\label{eq:R_exact_main}
\end{equation}
The cancellation is thus structural rather than the result of
fine-tuning: it follows from the weak \(U(1)\) symmetry and the
orthogonality between the population and slow-coherence sectors.
If several slow modes are degenerate, all contributing modes must be
removed to establish a change in the asymptotic decay rate; suppression
of one selected overlap alone is insufficient.

This behavior is evident in Fig.~\ref{fig:mpemba_regions}(c), where the
measurement-prepared dynamics originating from the pure initial state
exhibits a strongly suppressed long-time tail. The green and blue dashed curves show that mixed states also exhibit
strong finite suppression, as further reflected by the broad blue
region in Fig.~\ref{fig:mpemba_regions}(d). Away from
the excitation-conserving limit, \(\Omega_d\neq0\) mixes population and coherence
components of the Liouvillian eigenmodes, so that
\(\epsilon_c\) generally becomes finite and exact cancellation is no
longer guaranteed. Nevertheless, the steering protocol continues to
reduce the slow-mode overlap, as illustrated by the trajectories and
diagnostic region in Figs.~\ref{fig:mpemba_regions}(a) and
\ref{fig:mpemba_regions}(b). }

\begin{figure*}[t]
\centering
\hspace*{-1.6cm}    
\setlength{\abovecaptionskip}{-2pt}
\begin{overpic}[width=1\textwidth]{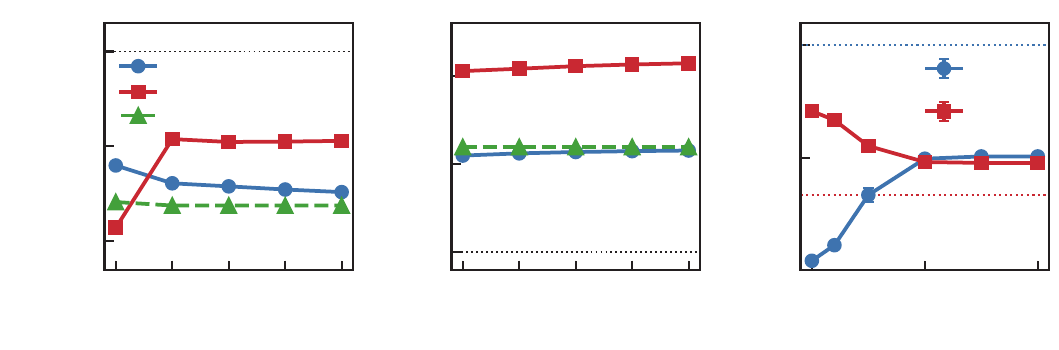}
\put(30.5,32){\scalebox{1.2}{(a)}}
\put(63.2,32){\scalebox{1.2}{(b)}}
\put(96,32){\scalebox{1.2}{(c)}}
\put(91.5,26.3){\scalebox{1}{$\langle|\mathcal R|\rangle$}}
\put(91.5,22){\scalebox{1}{$\langle\kappa_0\rangle/3$}}
\put(19.5,2.5){\scalebox{1.1}{$M$}}
\put(3.5,15.5){\rotatebox{90}{\scalebox{1.1}{$|\mathcal R|$}}}
\put(10.3,5.3){\scalebox{1.1}{$2$}}
\put(15.8,5.3){\scalebox{1.1}{$3$}}
\put(21.2,5.3){\scalebox{1.1}{$4$}}
\put(26.6,5.3){\scalebox{1.1}{$5$}}
\put(32,5.3){\scalebox{1.1}{$6$}}
\put(6.3,27.3){\scalebox{1.1}{$10^{0}$}}
\put(5.4,19){\scalebox{1.1}{$10^{-1}$}}
\put(5.4,9){\scalebox{1.1}{$10^{-2}$}}
\put(15.8,26.2){{\scalebox{1}{$\text{Driven, pure input}$}}}
\put(15.8,24){{\scalebox{1}{$\text{Driven, mixed input}$}}}
\put(15.8,21.8){{\scalebox{1}{$\text{Excitation conserving}$}}}
\put(43,5.3){\scalebox{1.1}{$2$}}
\put(48.5,5.3){\scalebox{1.1}{$3$}}
\put(54,5.3){\scalebox{1.1}{$4$}}
\put(59.4,5.3){\scalebox{1.1}{$5$}}
\put(64.8,5.3){\scalebox{1.1}{$6$}}
\put(40.7,25.3){\scalebox{1.1}{$3$}}
\put(40.7,17){\scalebox{1.1}{$2$}}
\put(40.7,8.5){\scalebox{1.1}{$1$}}
\put(53,2.5){\scalebox{1.1}{$M$}}
\put(38.5,16){\rotatebox{90}{\scalebox{1.1}{$\kappa_0$}}}
\put(75.6,5.3){\scalebox{1.1}{$0$}}
\put(86,5.3){\scalebox{1.1}{$0.1$}}
\put(97,5.3){\scalebox{1.1}{$0.2$}}
\put(87,2.5){\scalebox{1.2}{$\sigma_\theta$}}
\put(70,15.5){\rotatebox{90}{\scalebox{1.2}{$\text{criteria}$}}}
\put(73.7,28.3){\scalebox{1.1}{$1$}}
\put(72.5,17.5){\scalebox{1.1}{$0.5$}}
\put(73.7,6.8){\scalebox{1.1}{$0$}}
\end{overpic}
\vspace{0cm}
\caption{
Finite-size dependence and control-error robustness of the two
Mpemba diagnostics for the $M$-qubit model.
(a) Slow-mode amplitude ratio $|\mathcal R|$ and (b) initial-distance
ratio $\kappa_0$ vs qubit number $M$ for {driven local decay} with
pure (blue circles) and mixed (red squares) initial states, and for
the {excitation-conserving benchmark} (green triangles, pure input,
collective slow branch).
Dotted lines mark the thresholds $|\mathcal R|=1$ and $\kappa_0=1$.
(c) Mean $|\mathcal R|$ and $\kappa_0/3$ (scaled for display) versus
the standard deviation $\sigma_\theta$ of independent Gaussian errors
on the per-step measurement-basis ket angles $\theta_k$, for $M=3$
(driven case, pure initial state; 30 realizations, error bars show
one sample standard deviation).
Parameters as in Fig.~\ref{fig:mpemba_regions}.
}
\label{fss}
\end{figure*}

\section{Robustness and preparation cost}
\label{sec:robustness_cost}

\subsection{Finite-size dependence and robustness to control errors}
\label{app:finite_size}

We now examine how the overlap and initial-distance diagnostics behave as the Hilbert-space
dimension grows, and how they respond to imperfections in the measurement
axes. All parameters follow Fig.~\ref{fig:mpemba_regions}: $\Omega=1.2$,
$J=0.5$, $T=0.1$, $N=100$; the driven local-decay case uses
$(\gamma_1=1,\gamma_{j>1}=1.2,\Omega_d=0.1)$ with pure
$(|\mathbf r_1(0)|,\theta_1,\phi_1)=(1,\arccos(-0.6),0)$ and mixed
$(0.75,\arccos(-0.8),0)$ first-qubit initial states, and the excitation-conserving
case uses $(\gamma_j=1~\forall j,\,\Omega_d=0)$.

For these product initial states and uniform couplings, the Hamiltonian,
dissipative generator, and input state are invariant under permutations
of qubits $2,\ldots,M$. {A reduction to the permutation-invariant
operator sector is valid only if the measurement channels preserve that
sector as well. Under this condition, symmetrized Pauli strings reduce
the relevant operator-space dimension to $4\binom{M+2}{3}$ (e.g., $224$
for $M=6$ instead of $4^6=4096$). A general Gram--Schmidt completion of
the measurement basis need not satisfy this condition; in that case the
full prepared state and its subsequent dynamics must be retained.}

The size dependence of $|\mathcal R|$ and $\kappa_0$ is shown for
$M=2,\ldots,6$ [Figs.~\ref{fss}(a) and~\ref{fss}(b)]. Three features support the robustness of the mechanism over these
tested sizes.
First, in the driven local-decay case the suppressed slow-mode ratio does not
degrade over this size range: for the pure initial state $|\mathcal R|$
\emph{decreases} monotonically from $0.063$ ($M=2$) to $0.033$ ($M=6$),
while for the mixed state it saturates near $0.11$; both remain far below
unity. Second, $\kappa_0$ stays above unity and grows mildly with $M$
(from $2.10$ to $2.15$ for the pure state and from $3.03$ to $3.14$ for
the mixed state), so the prepared state remains initially farther from
stationarity. Third, the spectral gap does not close: $|{\rm
Re}\,\lambda_2|$ decreases only from $0.551$ to $0.517$ over the same
range. {A finite residual overlap reduces the amplitude of the slow
tail; exact removal of all contributing slow modes is required to
change the asymptotic decay exponent.} In the excitation-conserving benchmark, $|\mathcal R|$ is
set by the finite-$N$ off-diagonal residual of
Eq.~\eqref{eq:SM_comp_bound_simple} and is essentially independent of
$M$ ($|\mathcal R|\simeq0.0238$ for $M=3,\ldots,6$ at $N=100$), showing
that the finite suppression persists over the tested sizes.

Figure~\ref{fss}(c) probes control errors for $M=3$: each rotation angle
$\theta_k$ of the measurement basis acquires an independent Gaussian
perturbation of standard deviation $\sigma_\theta$, and $|\mathcal R|$
and $\kappa_0$ are averaged over $30$ noise realizations. The ensemble means satisfy both plotted
criteria over the tested noise range: even at
$\sigma_\theta=0.2\,{\rm rad}$ ($\approx11^\circ$ per step),
$\langle|\mathcal R|\rangle\simeq0.51<1$ and
$\langle\kappa_0\rangle\simeq1.43>1$. {These statements concern sampled means over the tested noise range,
not every realization or other control and readout errors.} Each
measurement still dephases the state in its chosen basis within
$\mathcal H_2$.

\subsection{State-side entropic cost}
\label{app:entropy_cost}

\begingroup

Final coherence in a fixed basis does not account for coherence
erased in successive measurement bases. Let $\rho_k^-$ and $\rho_k$
denote the states immediately before and after measurement $k$.
For mutually orthogonal projectors resolving the identity,
\begin{equation}
\rho_k
=
\sum_j P_j^{(k)}\rho_k^-P_j^{(k)},
\end{equation}
the entropy increase is
\begin{equation}
\Delta S_k
=
S(\rho_k)-S(\rho_k^-)
=
S(\rho_k^-\Vert\rho_k)
\geq 0,
\label{eq:entropy_increase_step}
\end{equation}
where 
$S(\rho\Vert\sigma)=\mathrm{Tr}[\rho(\ln\rho-\ln\sigma)]$.
For rank-1 measurements, $\Delta S_k$ is the relative-entropy
coherence erased in the instantaneous basis; higher-rank projectors
erase coherence between the resolved subspaces. Since the intervening
unitaries, including the final rotation, preserve entropy, the
cumulative mixing is
\begin{equation}
\Sigma_{\rm mix}
=
\sum_{k=1}^{N}\Delta S_k
=
S(\tilde\rho_N)-S(\rho_0).
\label{eq:Sigma_mix}
\end{equation}
This identity applies to arbitrary finite-dimensional inputs under
the assumed isolated preparation dynamics.
Its value depends on the physical measurement sequence, but is
invariant under a common change of representation of states and
projectors.

For the equal-step qubit sequence, the input and prepared Bloch lengths
are $r=|\mathbf r_0|$ and
$q_N=r\cos^N[\delta\theta_{\rm eff}/(N+1)]$, giving
\begin{equation}
\Sigma_{\rm mix}
=
h_2\!\left(\frac{1+q_N}{2}\right)
-
h_2\!\left(\frac{1+r}{2}\right),
\label{eq:qubit_mixing_cost}
\end{equation}
where $h_2(p)=-p\ln p-(1-p)\ln(1-p)$.
At fixed $0<r<1$ and fixed steering angle,
\begin{equation}
\Sigma_{\rm mix}
=
\frac{r\,\delta\theta_{\rm eff}^{\,2}}{4N}
\ln\frac{1+r}{1-r}
+O(N^{-2}).
\label{eq:mixed_cost_scaling}
\end{equation}
For a pure input,
$\varepsilon_N=(1-q_N)/2
=\delta\theta_{\rm eff}^{\,2}/(4N)+O(N^{-2})$, and
$\Sigma_{\rm mix}=h_2(\varepsilon_N)
=\varepsilon_N[1-\ln\varepsilon_N]+O(\varepsilon_N^2)$.
Thus, at fixed nonzero angle, the cost decreases as $1/N$ for
$0<r<1$ and as $(\ln N)/N$ for $r=1$, although the number of
measurements increases. The maximally mixed state is unchanged
and has $\Sigma_{\rm mix}=0$.

For the coupled-qubit example, we evaluate Eq.~(\ref{eq:Sigma_mix})
from the full $2^M$-dimensional density matrix. Although the basis
rotates only within $\mathcal H_2$, the first complete measurement
also erases coherences involving the complementary basis vectors.
For $M=6$ and the inputs and parameters of Fig.~\ref{fig:mpemba_regions},
$N=100$ gives $\Sigma_{\rm mix}=0.1118$ for the pure input and
$0.0978$ for the mixed input (natural logarithms), for both the
driven and undriven cases at the displayed precision. Increasing
$N$ from $10$ to $1000$ reduces these costs from $0.4548$ to
$0.0172$ and from $0.3980$ to $0.0151$, respectively.
Thus both examples quantify the coherence erased during preparation.
These multiqubit results use the complete prepared state; global
mixedness alone does not imply the mixed-qubit large-$N$ scaling.

The apparatus work, heat, and reset costs depend on the physical
implementation~\cite{Kam16}. A fixed effective Hamiltonian can still
require a laboratory drive, and adjustable measurement bases require
control operations. These costs cannot be inferred from
$\Sigma_{\rm mix}$, which is not the total entropy production of
the system and apparatus.
Appendix~\ref{app:fidelity_coherence} examines the tradeoff between
mixed-target fidelity and cumulative mixing.

\endgroup

\section{Experimental considerations}
\label{sec:experimental}

The protocol requires coherent
unitary control, nonselective projective measurements, and engineered
Markovian dissipation. These ingredients are available in trapped-ion
and superconducting platforms: calibrated laser, microwave, or
entangling-gate pulses for the unitary segments
\(U_k\)~\cite{Haf08,Kja20}; state-dependent fluorescence detection with
unconditional continuation (trapped ions)~\cite{Mye08,Cra19} or
dispersive mid-circuit readout (superconducting
circuits)~\cite{Jef14,Koh23} for the nonselective measurements; and
optical pumping or reservoir engineering (trapped
ions)~\cite{Lin13,Col22} or controlled qubit-environment coupling
(superconducting circuits)~\cite{Sha13,Kja20} for the dissipation.

{The duration \(T=\sum_{k=1}^{N+1}\Delta t_k\) counts only the
coherent intervals. For sequential operations, the full preparation time is}
{
\begin{equation}
T_{\rm prep}=T+\sum_{k=1}^{N}t_{{\rm meas},k}
+T_{\rm basis}+T_{\rm overhead},
\label{eq:prep_time}
\end{equation}
}
{including readout, basis changes, and additional device operations.
Neglecting dissipation requires this full duration to be short compared
with the relevant uncontrolled relaxation times, or requires the
dissipative coupling to be suppressed during preparation. For a
multiqubit state, a single-qubit value of $\gamma T_{\rm prep}$ alone
is not a bound on the full-state preparation error.}

As an illustrative timing estimate, take $T\gamma=0.1$ and an
engineered relaxation time $\gamma^{-1}=5\,{\rm ms}$. Then
\(T=0.5\,{\rm ms}\) and, for
\(N=100\), \(\Delta t_k\simeq5\,\mu{\rm s}\), longer than the quoted
superconducting readout times
\(t_{\rm meas}\sim0.1\)--\(1\,\mu{\rm s}\)~\cite{Jef14,Koh23};
trapped-ion readout with
\(t_{\rm meas}\sim10\)--\(11\,\mu{\rm s}\)~\cite{Cra19} has the
same ordering of readout and coherent-interval durations for
\(N\simeq20\)--\(30\) and \(\Delta t_k\) of order
\(20\,\mu{\rm s}\), while millisecond dissipative-state-preparation times have been
demonstrated~\cite{Lin13,Col22}. These reported preparation times
are not identified with the single rate $\gamma^{-1}$ assumed here.
At \(N\simeq20\)--\(30\) the steering remains effective, the geometric loss estimate \(\delta\theta_{\rm eff}^{2}/[2(N+1)]\) being of order
\(10^{-1}\) for the initial states considered here.

The measurement must preserve
populations in its chosen basis, for example through suitable
repeatable readout or ancilla-mediated dephasing; fast state
discrimination alone does not establish this channel. Any leakage
recovery or reset overhead must also be included. The relaxation curves
start after preparation. An overall time advantage to a chosen
trace-distance tolerance $\epsilon$ would additionally require
$T_{\rm prep}+t_\epsilon(\tilde\rho_N)<t_\epsilon(\rho_0)$,
where $t_\epsilon$ is the relaxation time to that tolerance.

\section{Conclusion}
\label{Conclusion}

We have developed a measurement-based framework for engineering the
quantum Mpemba effect in open quantum systems. A predissipative geometric
mode-steering stage reshapes the Liouvillian-mode decomposition of a given
state, while the subsequent Lindblad generator is kept fixed.
The protocol treats pure and mixed inputs, general Hamiltonians,
and real or complex slow spectral gaps. In limits where the slowest
modes are confined to the coherence sector, the measurement sequence can
approach removal of the entire slow sector, the condition for a strong
quantum Mpemba effect; the plotted slow-mode and initial-distance
diagnostics persist over the tested system sizes and measurement-axis
errors. {The state-side entropy increase quantifies cumulative
measurement-induced mixing and its tradeoff with mixed-target fidelity. Apparatus costs and total preparation-plus-relaxation times
require a specified implementation.}
 The protocol complements shortcuts to equilibration~\cite{Dan19}, Floquet-engineered thermalization control~\cite{Mar23}, and optimal-thermalization protocols~\cite{Kor22}. 

\begin{acknowledgments}
This work was supported by the National Natural Science Foundation of
China (Grant No.\ 12465009), the National Key R\&D Program of China
(Grant No.\ 2023YFA1406900), and the Major Program of the Jiangxi
Provincial Natural Science Foundation, China
(Grant No.\ 20224ACB201007).

\textit{Data availability}---There are no publicly available research data or software supporting this manuscript. Requests for further information or data should be sent to the authors.
\end{acknowledgments}
\clearpage
\onecolumngrid
\appendix

\section{Comparison with Related Slow-Mode-Suppression Protocols}
\label{app:comparison}

Table~\ref{tab:compare} summarizes how geometric mode steering relates to
representative protocols for controlling relaxation. All entries keep the physical comparison at the level
of (i) the control operations employed, (ii) the class of initial states
and spectral gaps covered, (iii) the knowledge and resources required, and (iv) whether the relaxation generator is modified by the protocol.

\begin{table}[h]
\centering
\caption{Comparison of {representative relaxation-control} protocols.
{The state and spectral assumptions describe the cited constructions,
not restrictions on every implementation of each control operation.}
``Generator fixed'' indicates {whether the relaxation dynamics are
left unchanged as part of the construction.}}
\label{tab:compare}
\renewcommand{\arraystretch}{1.8}
\setlength{\tabcolsep}{7pt}
\begin{tabular}{lllll}
\hline\hline
\tcell{Protocol} &
\tcell{Control operations} &
\tcell{States / gaps} &
\tcell{Required knowledge} &
\tcell[2.4cm]{Generator fixed} \\
\hline
\tcell{Mode-designed unitary~\cite{Car21}} &
\tcell{one designed unitary} &
\tcell{pure states; isolated real slow mode} &
\tcell{slow left eigenoperator and initial state} &
\tcell[2.4cm]{yes} \\[3pt]

\tcell{Collective spin rotations~\cite{Koc22}} &
\tcell{global rotations about two spin axes} &
\tcell{pure product inputs; real or complex slow modes} &
\tcell{initial state and dynamics; optimization of rotation angles} &
\tcell[2.4cm]{yes} \\[3pt]

\tcell{Davies-map diagonalization~\cite{Mor24}} &
\tcell{diagonalizing unitary (+ population inversion)} &
\tcell{pure or mixed states; Davies maps} &
\tcell{full eigenbasis of $\rho_0$; energy eigenbasis} &
\tcell[2.4cm]{yes} \\[3pt]

\tcell{Temporary reset~\cite{Bao25}} &
\tcell{transient coupling to an auxiliary reset channel} &
\tcell{reset-assisted relaxation} &
\tcell{reset target; access to the auxiliary channel} &
\tcell[2.4cm]{no (transiently modified)} \\[3pt]

\tcell{Continuous control~\cite{Dan19}} &
\tcell{time-dependent driving during relaxation} &
\tcell{thermal-state transfer; driven oscillator} &
\tcell{controlled Hamiltonian and dissipative model} &
\tcell[2.4cm]{no (driven during relaxation)} \\[3pt]

\tcell{Geometric mode steering (this work)} &
\tcell{free evolutions under fixed $H_{\rm eff}$ interleaved with nonselective measurements} &
\tcell{pure and mixed states; real and complex gaps} &
\tcell{relevant slow modes, input direction or steering subspace, and target} &
\tcell[2.4cm]{yes} \\
\hline\hline
\end{tabular}
\end{table}

{The coherent segments are free evolutions under fixed $H_{\rm eff}$;
the measurement bases provide the control freedom. A qubit schedule
requires the input Bloch direction, whereas Davies-map diagonalization
uses the full state eigenbasis~\cite{Mor24}. The multiqubit construction
requires a selected dominant input eigenstate, a target, and an
implementable basis for the steering subspace. These requirements do
not imply lower information or gate costs. Measurements erase coherence
in their successive bases; whether this suppresses slow relaxation must
be checked from the overlaps of the full prepared state.}

\section{Trace Distance versus Relative Entropy in the Geometric-Mode-Steering Quantum Mpemba Effect}
\label{sec:trace_vs_relative}
 
In the main text, we employ the trace distance to quantify the distance to
stationary states. Here we justify this choice by showing that the trace
distance provides a direct diagnostic of slow-mode suppression, which is the
central mechanism underlying the geometric-mode-steering-induced quantum
Mpemba effect.
 
We consider an open quantum system governed by a time-independent, diagonalizable Lindblad
generator $\mathcal{L}$ with stationary state $\rho_{\mathrm{ss}}$.
The state is expanded in Liouvillian eigenmodes as
\begin{equation}
\rho_t = \rho_{\mathrm{ss}}
       + \sum_{\nu\ge 2} \mathrm{Tr}(l_\nu \rho_0)\,r_\nu\,e^{\lambda_\nu t},
\label{eq:eigen_expand}
\end{equation}
where $\lambda_\nu$ are the eigenvalues of $\mathcal{L}$ ordered by
increasing $|\mathrm{Re}(\lambda_\nu)|$, $r_\nu$ are the corresponding
right eigenoperators, and $l_\nu$ are the left eigenoperators satisfying
the biorthogonality condition $\mathrm{Tr}(l_\nu r_\mu)=\delta_{\nu\mu}$.
The single-mode asymptotic formulas below assume a single nondegenerate
real slowest mode with nonzero overlap.
 
\textit{Trace distance.}---The trace distance~\cite{Nie10,Car21} is defined
as
\begin{equation}
D_{\mathrm{T}}(\rho_t, \rho_{\mathrm{ss}})
= \frac{1}{2}\|\rho_t-\rho_{\mathrm{ss}}\|_1
= \frac{1}{2}
  \left\|\sum_{\nu\ge 2}
  \mathrm{Tr}(l_\nu\rho_0)\,r_\nu\,e^{\lambda_\nu t}
  \right\|_1.
\end{equation}
At long times, it is dominated by the slowest mode $\nu=2$,
\begin{equation}
D_{\mathrm{T}}(\rho_t, \rho_{\mathrm{ss}})
\;\sim\;
\big|\mathrm{Tr}(l_2\rho_0)\big|\,
\tfrac{1}{2}\|r_2\|_1\,
e^{\mathrm{Re}(\lambda_2)t}.
\label{eq:DT_slowmode}
\end{equation}
Since $\tfrac{1}{2}\|r_2\|_1$ is a mode-dependent constant fixed by the
normalization of $r_2$, Eq.~\eqref{eq:DT_slowmode} shows that
$D_{\mathrm{T}}$ depends \emph{linearly} on the slow-mode overlap
$|\mathrm{Tr}(l_2\rho_0)|$.
When the slowest eigenvalue belongs to a complex-conjugate
pair $(\lambda_2,\lambda_2^*)$, the sum in
Eq.~\eqref{eq:eigen_expand} contains both members, and the linear scaling in
Eq.~\eqref{eq:DT_slowmode} applies to the \emph{envelope} of the
long-time tail: since
$\mathrm{Tr}(l_2^\dagger\rho_0)=[\mathrm{Tr}(l_2\rho_0)]^*$ for Hermitian
$\rho_0$, the two members carry identical amplitude moduli, and the tail
decays as $|\mathrm{Tr}(l_2\rho_0)|\,e^{\mathrm{Re}(\lambda_2)t}$
multiplied by a bounded oscillatory factor that can depend on the overlap
phase. The linear dependence on the single overlap modulus is unchanged
at the envelope level.
Any protocol that reduces this overlap therefore produces a directly
proportional suppression of the long-time relaxation amplitude or envelope.
In the geometric mode-steering protocol, repeated measurements transform $\rho_0$ into
$\tilde{\rho}_N$ so that
$|\mathrm{Tr}(l_2\tilde{\rho}_N)|\ll|\mathrm{Tr}(l_2\rho_0)|$,
and Eq.~\eqref{eq:DT_slowmode} guarantees that this suppression is
asymptotically and proportionally reflected in the amplitude or envelope of
$D_{\mathrm{T}}(\rho_t, \rho_{\mathrm{ss}})$.
 
\textit{Quantum relative entropy.}---The quantum relative
entropy~\cite{Mor24} is defined as
\begin{equation}
S(\rho_t\|\rho_{\mathrm{ss}})
= \mathrm{Tr}\!\left[\rho_t(\ln\rho_t - \ln\rho_{\mathrm{ss}})\right].
\end{equation}
Writing $\rho_t = \rho_{\mathrm{ss}}+\delta\rho_t$ with
$\mathrm{Tr}(\delta\rho_t)=0$ and using the Fr\'{e}chet expansion of the
operator logarithm, the relative entropy admits the quadratic expansion
\begin{equation}
S(\rho_t\|\rho_{\mathrm{ss}})
= \frac{1}{2}\,
  \mathrm{Tr}\!\left[
  \delta\rho_t\,\mathcal{K}^{-1}_{\rho_{\mathrm{ss}}}(\delta\rho_t)
  \right]
  + O(\|\delta\rho_t\|^3),
\label{eq:QRE_expand}
\end{equation}
where the inverse Kubo--Mori (Bogoliubov) operator is
\begin{equation}
\mathcal{K}^{-1}_{\rho_{\mathrm{ss}}}(A)
= \int_0^\infty ds\,
  (\rho_{\mathrm{ss}}+s\,I)^{-1}A\,(\rho_{\mathrm{ss}}+s\,I)^{-1},
\end{equation}
with $I$ the identity operator and $A$ a traceless Hermitian operator.
Since $\delta\rho_t \propto \mathrm{Tr}(l_2\rho_0)\,r_2\,e^{\lambda_2 t}$
at long times, Eq.~\eqref{eq:QRE_expand} gives
\begin{equation}
S(\rho_t\|\rho_{\mathrm{ss}})
\;\propto\;
\big|\mathrm{Tr}(l_2\rho_0)\big|^2\,
e^{2\,\mathrm{Re}(\lambda_2)t}.
\label{eq:QRE_slowmode}
\end{equation}
The relative entropy therefore depends \emph{quadratically} on the
slow-mode overlap, in contrast to the linear dependence of the trace distance in
Eq.~\eqref{eq:DT_slowmode}.
 
\textit{Implications for the quantum Mpemba effect.}---Equations
\eqref{eq:DT_slowmode} and \eqref{eq:QRE_slowmode} make this
distinction explicit.
When the geometric mode-steering protocol reduces the slow-mode amplitude by a factor
$0<|\mathcal{R}| = |\mathrm{Tr}(l_2\tilde{\rho}_N)|/|\mathrm{Tr}(l_2\rho_0)|
< 1$, the long-time behavior of the two measures scales as
\begin{align}
D_{\mathrm{T}}(\tilde\rho_t, \rho_{\mathrm{ss}})
&\;\sim\; |\mathcal{R}|\;
  D_{\mathrm{T}}(\rho_t, \rho_{\mathrm{ss}}),
\label{eq:DT_ratio}\\
S(\tilde\rho_t\|\rho_{\mathrm{ss}})
&\;\sim\; |\mathcal{R}|^2\;
  S(\rho_t\|\rho_{\mathrm{ss}}).
\label{eq:QRE_ratio}
\end{align}
For a complex-conjugate pair, these relations describe envelope scalings;
same-time proportionality additionally requires identical normalized
oscillatory factors.
The suppression of the slow mode is reflected \emph{linearly} in the trace
distance and \emph{quadratically} in the relative entropy.
The trace distance therefore provides a direct, linear measure of
the slow-mode amplitude controlled by the protocol.
This is the primary reason for adopting the trace distance in the main text.
 
\textit{Role of coherences.}---We further illustrate the comparison by
examining how each measure responds to population and coherence
deviations in a single-qubit example.
Consider
\begin{equation}
\rho_t =
\begin{pmatrix} p & \alpha \\ \alpha^\ast & 1-p \end{pmatrix},
\qquad
\rho_{\rm ss} =
\begin{pmatrix} p_{\rm ss} & 0 \\ 0 & 1-p_{\rm ss} \end{pmatrix},
\end{equation}
with $0<p_{\rm ss}<1$ and $\delta p = p-p_{\rm ss}$.
The difference matrix $\rho_t - \rho_{\rm ss}$ has eigenvalues
$\pm\sqrt{(\delta p)^2+|\alpha|^2}$, so
\begin{equation}
D_{\rm T}(\rho_t,\rho_{\rm ss})
= \sqrt{(\delta p)^2+|\alpha|^2}.
\label{eq:DT_qubit}
\end{equation}
For the same state, the leading-order relative entropy is
\begin{equation}
S(\rho_{\rm ss}+\delta\rho\,\|\,\rho_{\rm ss})
= \frac{(\delta p)^2}{2p_{\rm ss}(1-p_{\rm ss})}
+ \frac{\ln[p_{\rm ss}/(1-p_{\rm ss})]}{2p_{\rm ss}-1}\,|\alpha|^2
+ O(\|\delta\rho\|^3).
\label{eq:QRE_qubit}
\end{equation}
Equation~\eqref{eq:QRE_qubit} shows that both the population deviation
$\delta p$ and the coherence amplitude $|\alpha|$ enter the relative entropy
at \emph{second order}.
By contrast, Eq.~\eqref{eq:DT_qubit} shows that the trace distance treats
population and coherence deviations on an equal footing: both $\delta p$ and
$|\alpha|$ appear under the same square root, so neither is systematically
suppressed relative to the other.
This behavior is consistent with the slow-mode analysis above.
The long-time deviation $\rho_t - \rho_{\rm ss} \simeq
\mathrm{Tr}(l_2\rho_0)\,r_2\,e^{\lambda_2 t}$ generically contains both
population and coherence components in $r_2$.
The trace distance responds to the full deviation through the single linear
factor $|\mathrm{Tr}(l_2\rho_0)|$ in Eq.~\eqref{eq:DT_slowmode}, capturing
both components simultaneously, whereas the relative entropy depends on the
entire deviation quadratically through Eq.~\eqref{eq:QRE_slowmode}.
 
In summary, the trace distance is preferred over the relative entropy
because (i)~it depends \emph{linearly} on the slow-mode overlap
[Eq.~\eqref{eq:DT_slowmode}], so that the slow-mode suppression
achieved by the geometric mode-steering protocol is reflected directly and proportionally
in the long-time relaxation amplitude or envelope; and (ii)~it responds to population
and coherence deviations on an equal footing [Eq.~\eqref{eq:DT_qubit}],
consistently capturing the full structure of the deviation
$\rho_t - \rho_{\rm ss}$.

Two further remarks are in order. First, the quadratic (Kubo--Mori)
expansion \eqref{eq:QRE_expand} presupposes a full-rank stationary
state. In the undriven, zero-temperature qubit limit, the steady state
becomes pure, $\rho_{\rm ss}=|g\rangle\langle g|$, and
$S(\rho_t\|\rho_{\rm ss})$ diverges for any state with support outside
$|g\rangle$, whereas the trace distance remains well defined and
bounded, $0\le D_{\rm T}\le 1$. The Davies-type benchmark discussed in
the main text is taken precisely in this limit, which provides an
additional practical reason for adopting $D_{\rm T}$ as the distance
measure. Relatedly, the nonequilibrium free energy employed for Davies
maps in Ref.~\cite{Mor24}, $F_{\rm neq}[\rho]=\beta^{-1}S(\rho\|\tau_\beta)
+F_{\rm eq}$, presupposes a \emph{thermal} fixed point $\tau_\beta$ at
inverse temperature $\beta$; for the driven and local-dissipator
Liouvillians considered in the main text the stationary state need not be
of Gibbs form for the system Hamiltonian at a bath temperature, so the
above free-energy interpretation of $S(\rho\|\rho_{\rm ss})$ does not
generally apply, whereas $D_{\rm T}$ requires only a specified stationary state. Second, the coefficient
$\ln[p_{\rm ss}/(1-p_{\rm ss})]/(2p_{\rm ss}-1)$ multiplying
$|\alpha|^2$ in Eq.~\eqref{eq:QRE_qubit} has a removable singularity at
$p_{\rm ss}=1/2$, with limiting value $2$, equal to the population
coefficient $1/[2p_{\rm ss}(1-p_{\rm ss})]$ evaluated at the same
point; populations and coherences thus contribute with equal weight at
the maximally mixed stationary state, as required by symmetry.

\section{Derivation of the Relaxation-Acceleration Condition}\label{app:accel_condition}

In this Appendix, we derive the slow-mode suppression condition used to
identify relaxation acceleration in the driven dissipative qubit.
The system dynamics is governed by the Lindblad master equation \cite{Zha22}
\begin{equation}
\dot{\rho} = -i[H,\rho] + \frac{\gamma}{2}
\left( 2\sigma^-\rho\sigma^+ - \sigma^+\sigma^-\rho
- \rho\sigma^+\sigma^- \right)
\equiv \mathcal{L}\rho,
\label{eq:SM_L}
\end{equation}
where \(H\) is the Hamiltonian defined in Eq.~\eqref{eq:qubit_H}.
Here $\gamma$ is the spontaneous emission rate, $\sigma^+=\ket{e}\bra{g}$ and $\sigma^-=\ket{g}\bra{e}$ are the raising and lowering operators.

To diagonalize the Liouvillian $\mathcal{L}$, we vectorize the density matrix according to
\begin{equation}
\rho \;\longrightarrow\;
\ket{\rho} =
\begin{pmatrix}
\rho_{ee} \\
\rho_{eg} \\
\rho_{ge} \\
\rho_{gg}
\end{pmatrix},
\end{equation}
so that $\mathcal{L}$ acts as a $4\times4$ matrix. Substituting
Eqs.~\eqref{eq:qubit_H} and \eqref{eq:SM_L}  yields
\begin{equation}
\mathcal{L} =
\begin{pmatrix}
-\gamma & i\Omega/2 & -i\Omega/2 & 0 \\
i\Omega/2 & -i\Delta - \gamma/2 & 0 & -i\Omega/2 \\
-i\Omega/2 & 0 & i\Delta - \gamma/2 & i\Omega/2 \\
\gamma & -i\Omega/2 & i\Omega/2 & 0
\end{pmatrix}.
\label{eq:SM_Lmat}
\end{equation}

Since $\mathcal{L}$ preserves the trace, it has a zero eigenvalue
$\lambda_1=0$ corresponding to the steady state. The associated left
eigenvector is
\begin{equation}
l_1 = (1,\,0,\,0,\,1),
\end{equation}
which implements the trace operation,
$l_1\ket{\rho}=\mathrm{Tr}(\rho)$.
The corresponding right eigenvector satisfies
$\mathcal{L}[r_1]=0$, yielding
\begin{equation}
r_1 =
\frac{1}{N_1}
\begin{pmatrix}
\Omega^2 \\
-\Omega(2\Delta + i\gamma) \\
-\Omega(2\Delta - i\gamma) \\
4\Delta^2 + \Omega^2 + \gamma^2
\end{pmatrix},
\label{eq:SM_PhiR1}
\end{equation}
with normalization fixed by $l_1r_1=1$, giving
$N_1 = 4\Delta^2 + 2\Omega^2 + \gamma^2$.
Reshaping $r_1$ gives the steady-state density matrix
\begin{equation}
\rho_{\mathrm{ss}} =
\frac{1}{4\Delta^2 + 2\Omega^2 + \gamma^2}
\begin{pmatrix}
\Omega^2 & -\Omega(2\Delta + i\gamma) \\
-\Omega(2\Delta - i\gamma) & 4\Delta^2 + \Omega^2 + \gamma^2
\end{pmatrix}.
\end{equation}

We now consider the nonzero eigenvalues $\lambda_\nu$ ($\nu=2,3,4$).
For any decaying mode, trace preservation implies a tracelessness
constraint. Applying $l_1$ to
$\mathcal{L}[r_\nu]=\lambda_\nu r_\nu$ gives
\begin{equation}
l_1 r_\nu = x_1 + x_4 = 0,
\label{eq:trace_constraint}
\end{equation}
where $ r_\nu=(x_1,x_2,x_3,x_4)^T$.
At generic parameter points where the following denominators are
nonzero, we may choose $x_1=1$ and $x_4=-1$. The remaining components are
\begin{equation}
x_2 = \frac{i\Omega}{\gamma/2 + i\Delta + \lambda_\nu},
\qquad
x_3 = \frac{i\Omega}{-\gamma/2 + i\Delta - \lambda_\nu}.
\label{eq:x2x3}
\end{equation}
Thus,
\begin{equation}
 r_\nu = \frac{1}{N_\nu}
\begin{pmatrix}
1 \\
\dfrac{i\Omega}{\gamma/2 + i\Delta + \lambda_\nu} \\[6pt]
\dfrac{i\Omega}{-\gamma/2 + i\Delta - \lambda_\nu} \\[6pt]
-1
\end{pmatrix}.
\label{eq:SM_PhiRnu}
\end{equation}

The trace-dual left eigenoperators satisfy
$\mathcal{L}^\dagger[l_\nu]=\lambda_\nu l_\nu$, so that
$\mathrm{Tr}(l_\nu\rho_t)=e^{\lambda_\nu t}\mathrm{Tr}(l_\nu\rho_0)$.
The following row lists the coefficients of this functional in the
chosen vectorization (its middle entries are $(l_\nu)_{ge}$ and
$(l_\nu)_{eg}$). Solving componentwise gives
\begin{equation}
l_\nu =
\begin{pmatrix}
1 &
\dfrac{-2i\Omega \lambda_\nu}{(\gamma - \lambda_\nu)(\gamma + 2i\Delta
+ 2\lambda_\nu)} &
\dfrac{-2i\Omega \lambda_\nu}{(\gamma - \lambda_\nu)(-\gamma + 2i\Delta
- 2\lambda_\nu)} &
\dfrac{\lambda_\nu + \gamma}{\gamma - \lambda_\nu}
\end{pmatrix}.
\label{eq:PhiL}
\end{equation}

Away from defective points, the eigenvectors of $\mathcal{L}$ form a
biorthogonal set, $\mathrm{Tr}(l_\nu r_\mu)=\delta_{\nu \mu}$, and
provide a complete basis in Liouville space. Biorthogonality leaves the
rescaling freedom $l_\nu\to\xi_\nu l_\nu$, $r_\nu\to r_\nu/\xi_\nu$
with $\xi_\nu\in\mathbb{C}$. For the decaying modes $\nu\ge2$, we fix the modulus of this
rescaling by requiring $\|l_\nu\|_F=1$, where $\|A\|_F=\sqrt{\mathrm{Tr}(A^\dagger A)}$
denotes the Frobenius norm.  The ratio $\mathcal{R}$ defined
in Eq.~\eqref{rr0} of the main text is independent of $\xi_\nu$---both modulus
and phase cancel between numerator and denominator---whereas the
absolute modal weights $|\mathrm{Tr}(l_\nu\rho)|$ depend on this convention.

The Liouvillian has a unique steady state, while all other eigenvalues
have negative real parts. Let $\lambda_2$ denote the eigenvalue with the
largest real part among the nonzero modes. The corresponding left
eigenoperator $l_2$, normalized as above, can be expanded in the Pauli basis as
\begin{equation}
l_2 = \beta_0 I_2 + \boldsymbol{\beta}\cdot\boldsymbol{\sigma},
\label{eq:SM_l2_expand}
\end{equation}
where $\boldsymbol{\beta}=(\beta_x,\beta_y,\beta_z)$.
Matching components with Eq.~\eqref{eq:PhiL} yields
\begin{equation}
\beta_0 = \frac{\gamma \xi_2}{\gamma-\lambda_2},
\label{eq:SM_beta0}
\end{equation}
and
\begin{equation}
\boldsymbol{\beta}
=
\frac{-\lambda_2 \xi_2}{\gamma-\lambda_2}
\begin{pmatrix}
\dfrac{4\Omega\Delta}{(\gamma+2\lambda_2)^2+4\Delta^2} \\[10pt]
\dfrac{2\Omega(\gamma+2\lambda_2)}{(\gamma+2\lambda_2)^2+4\Delta^2} \\[10pt]
1
\end{pmatrix}.
\label{eq:SM_beta_vec}
\end{equation}

For a general Bloch state
$\rho = \frac12\left(I_2+\mathbf r\cdot\boldsymbol{\sigma}\right)$,
we obtain
\begin{equation}
\mathrm{Tr}(l_2\rho)
=
\beta_0 + \boldsymbol{\beta}\cdot\mathbf r.
\label{eq:SM_overlap}
\end{equation}
This quantity measures the weight of the slowest decaying mode.
The explicit parametrization assumes nonzero denominators and
$\beta_z\ne0$. {At singular parameter points the eigenoperators must be
obtained directly; ties between leading decay rates require the complete
slow sector rather than one selected mode.}

We now consider the optimized nonselective measurement-steering protocol. In the frequent-measurement limit, the Bloch vector is
projected onto the $\hat{z}$ axis,
\begin{equation}
\lim_{N\to\infty}\tilde{\mathbf r}_N = (0,0,|\mathbf r_0|)^\top.
\label{eq:SM_rN}
\end{equation}
Using Eq.~\eqref{eq:SM_overlap}, we define
\begin{equation}
\mathcal{R}_\infty
=
\lim_{N\to\infty}\frac{\mathrm{Tr}(l_2\tilde{\rho}_N)}
{\mathrm{Tr}(l_2\rho_0)}
=
\frac{\beta_0 + |\mathbf r_0|\beta_z}
{\beta_0 + \boldsymbol{\beta}\cdot\mathbf r_0}.
\label{eq:SM_R}
\end{equation}

For $\beta_z\neq0$ we introduce the ratios
\begin{equation}
a=\frac{\beta_0}{|\mathbf r_0|\beta_z},
\qquad
b_x=\frac{\beta_x}{\beta_z},
\qquad
b_y=\frac{\beta_y}{\beta_z},
\label{eq:SM_abxy_def}
\end{equation}
which are independent of the normalization factor $\xi_2$ and, by
Eqs.~\eqref{eq:SM_beta0} and \eqref{eq:SM_beta_vec}, evaluate to
\begin{equation}
a=-\frac{\gamma}{|\mathbf r_0|\lambda_2},
\quad
b_x=\frac{4\Omega\Delta}{D},
\quad
b_y=\frac{2\Omega(\gamma+2\lambda_2)}{D},
\label{eq:SM_abxy_val}
\end{equation}
with $D=(\gamma+2\lambda_2)^2+4\Delta^2$. Parametrizing
$\mathbf r_0=|\mathbf r_0|
(\sin\theta\cos\phi,\sin\theta\sin\phi,\cos\theta)$
and introducing the (generally complex) projection
\begin{equation}
X
\equiv
b_x\sin\theta\cos\phi
+
b_y\sin\theta\sin\phi
+
\cos\theta,
\label{eq:SM_X_def}
\end{equation}
the limiting slow-mode amplitude ratio \eqref{eq:SM_R} takes the compact form
\begin{equation}
\mathcal R_\infty
=
\frac{a+1}{a+X}.
\label{eq:SM_R_compact}
\end{equation}
For completeness, define $\mathbf u_N=\tilde{\mathbf r}_N/|\mathbf r_0|$
using Eq.~\eqref{eq:finiteN_qubit}. The exact finite-measurement result is
\begin{equation}
\mathcal R_N=
\frac{\beta_0+|\mathbf r_0|\boldsymbol\beta\cdot\mathbf u_N}
{\beta_0+\boldsymbol\beta\cdot\mathbf r_0}
=\frac{a+b_xu_{Nx}+b_yu_{Ny}+u_{Nz}}{a+X}.
\label{eq:SM_R_finite}
\end{equation}
{Thus the residual tilt and contraction both enter the finite-$N$ ratio.
A vanishing denominator denotes zero original overlap and requires a
direct comparison, rather than assigning a ratio.}

\begin{figure}[htb]
\centering
\hspace*{0.55cm}
\begin{overpic}[width=5.57cm]
{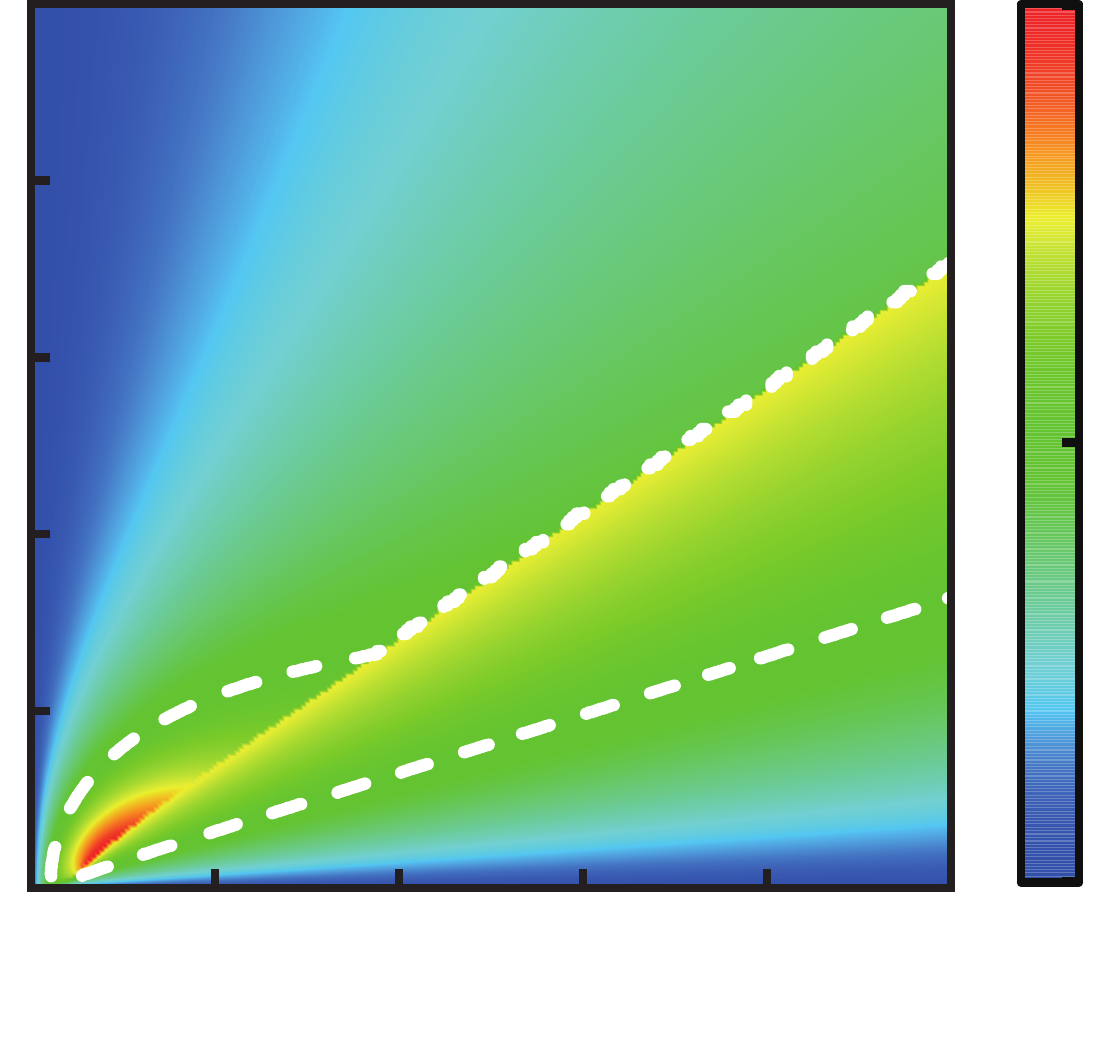}
\put(78,99){\scalebox{1.1}{(a)}}
\put(90,98){\scalebox{1}{$|\mathcal R|$}}
\put(45,2){\scalebox{1}{$\Omega$}}
\put(45,2){\scalebox{1}{$\Omega$}}
\put(-12,50){\rotatebox{90}{\scalebox{1}{$\Delta$}}}
\put(-4,92){\scalebox{1.0}{5}}
\put(-4,77){\scalebox{1.0}{4}}
\put(-4,62){\scalebox{1.0}{3}}
\put(-4,45){\scalebox{1}{2}}
\put(-4,29){\scalebox{1}{1}}
\put(-4,15){\scalebox{1}{0}}
\put(1,8){\scalebox{1}{0}}
\put(18,8){\scalebox{1}{$1$}}
\put(34,8){\scalebox{1}{$2$}}
\put(51,8){\scalebox{1}{$3$}}
\put(68,8){\scalebox{1}{$4$}}
\put(84,8){\scalebox{1}{$5$}}
\put(100,15){\scalebox{0.9}{$0$}}
\put(100,55){\scalebox{0.9}{$1$}}
\put(100,94){\scalebox{0.9}{$1.9$}}
\end{overpic}
\hspace{0.35in}
\hspace*{-0.2cm}
\begin{overpic}[width=5.57cm]
{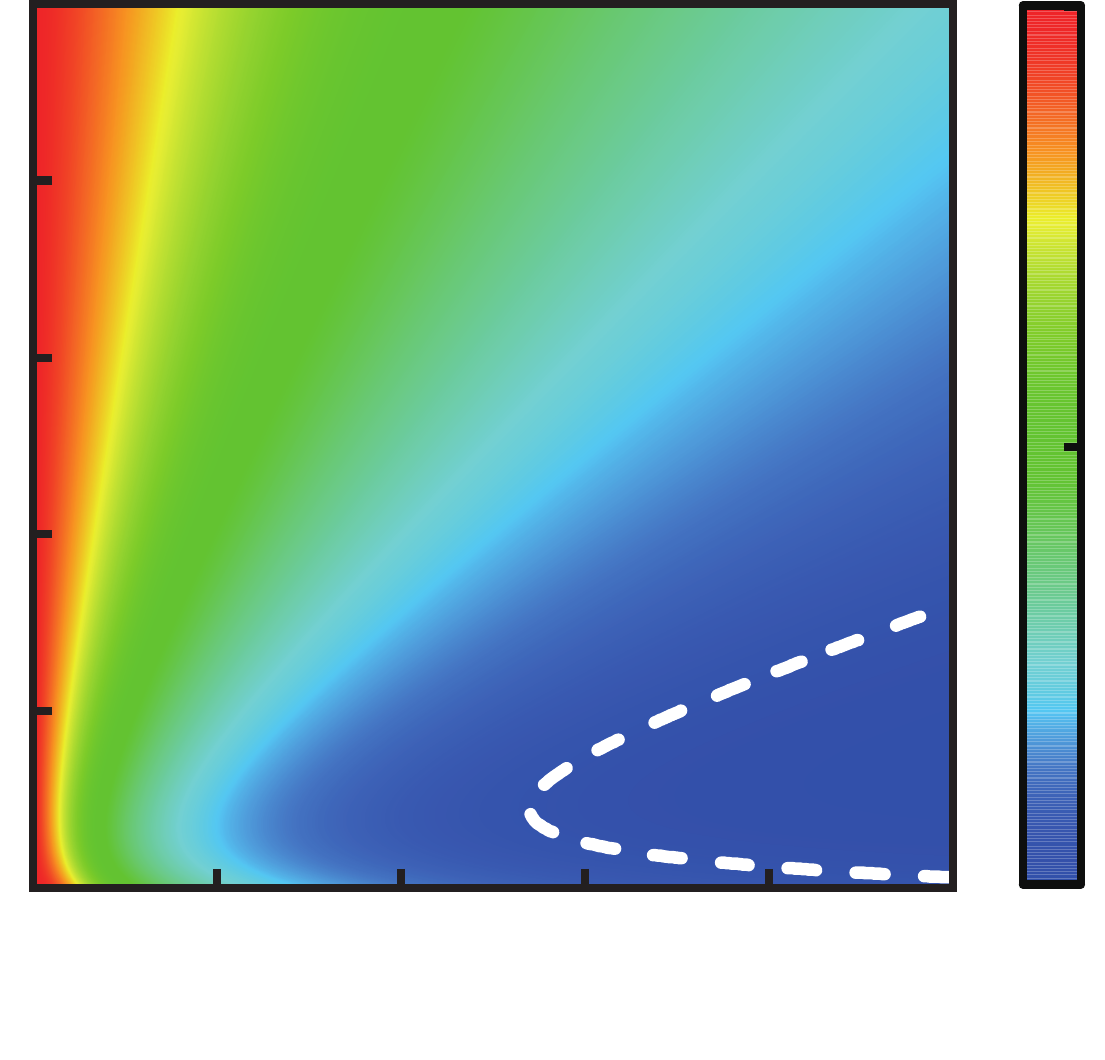}
\put(78,99){\scalebox{1.1}{(b)}}
\put(-13,50){\rotatebox{90}{\scalebox{1.0}{$\Delta$}}}
\put(93,98){\scalebox{1.0}{$\kappa_0$}}
\put(45,2){\scalebox{1}{$\Omega$}}

\put(-4,92){\scalebox{1.0}{5}}
\put(-4,77){\scalebox{1.0}{4}}
\put(-4,62){\scalebox{1.0}{3}}
\put(-4,45){\scalebox{1}{2}}
\put(-4,29){\scalebox{1}{1}}
\put(-4,15){\scalebox{1}{0}}
\put(1,8){\scalebox{1}{0}}
\put(18,8){\scalebox{1}{$1$}}
\put(35,8){\scalebox{1}{$2$}}
\put(52,8){\scalebox{1}{$3$}}
\put(68,8){\scalebox{1}{$4$}}
\put(84,8){\scalebox{1}{$5$}}
\put(100,15){\scalebox{0.9}{$0.9$}}
\put(100,55){\scalebox{0.9}{$1.9$}}
\put(100,94){\scalebox{0.9}{$2.9$}}
\end{overpic}
\par\vspace*{0.2cm}
\hspace*{-0.05cm}%
\begin{overpic}[width=5.57cm]
{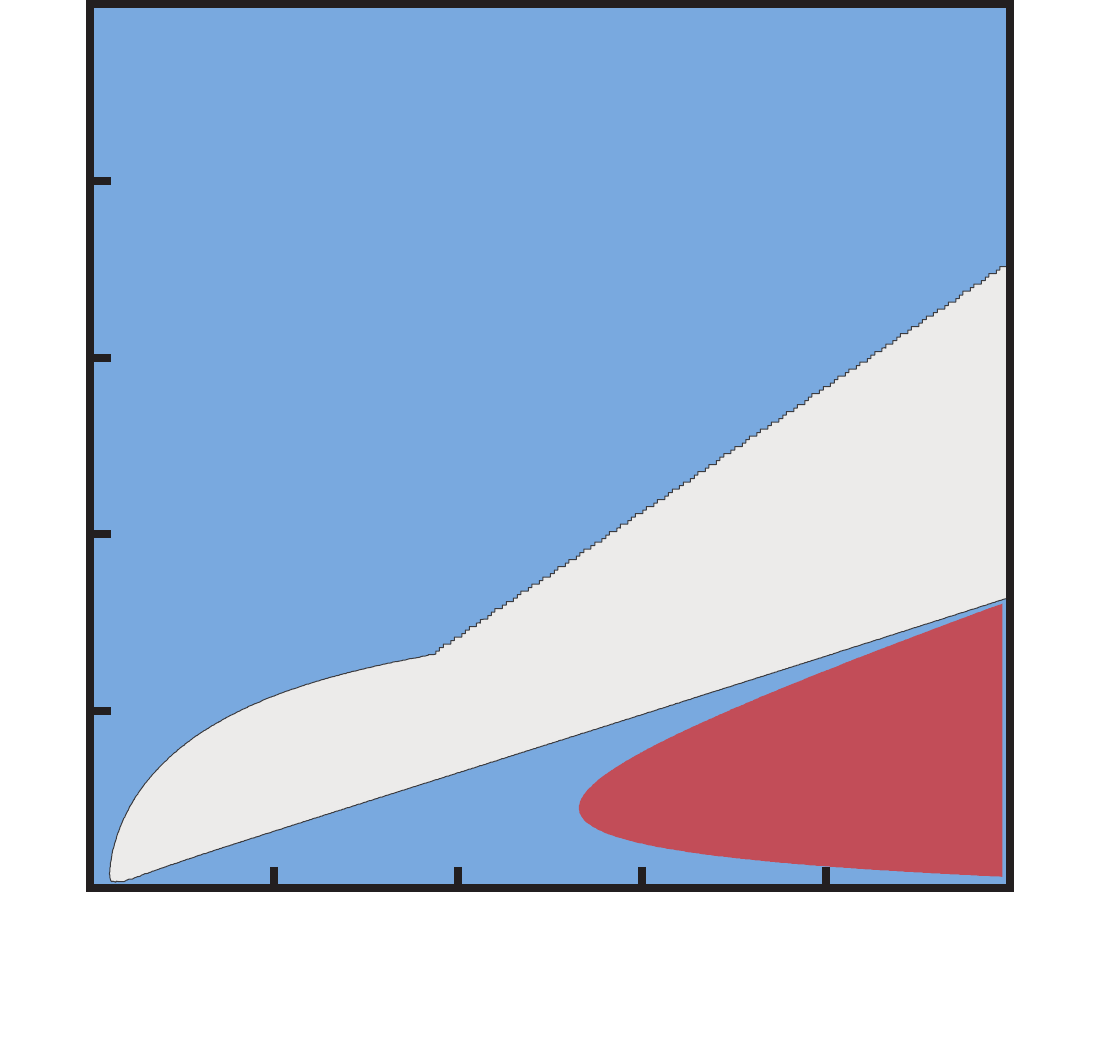}
\put(82,99){\scalebox{1.1}{(c)}}
\put(29,69){\scalebox{0.8}{\shortstack{\textbf{Mpemba}\\ ($|\mathcal{R}|<1,\ \kappa_0>1$)}}}
\put(63,27){\scalebox{0.8}{acceleration}}
\put(54,22){\scalebox{0.7}{($|\mathcal{R}|<1,\ \kappa_0 \le 1$)}}
\put(50,35){\rotatebox{31}{\scalebox{0.8}{\shortstack{\\ ($|\mathcal{R}|\ge1,\ \kappa_0>1$)}}}}

\put(-5,50){\rotatebox{90}{\scalebox{1.0}{$\Delta$}}}
\put(48,2){\scalebox{1}{$\Omega$}}
\put(2,92){\scalebox{1.0}{5}}
\put(2,77){\scalebox{1.0}{4}}
\put(2,62){\scalebox{1.0}{3}}
\put(2,45){\scalebox{1}{2}}
\put(2,29){\scalebox{1}{1}}
\put(2,15){\scalebox{1}{0}}
\put(7,8){\scalebox{1}{0}}
\put(23.5,8){\scalebox{1}{$1$}}
\put(40.2,8){\scalebox{1}{$2$}}
\put(56.8,8){\scalebox{1}{$3$}}
\put(73.3,8){\scalebox{1}{$4$}}
\put(89.4,8){\scalebox{1}{$5$}}
\end{overpic}
\hspace{0.35in}
\hspace*{-0.2cm}
\begin{overpic}[width=5.57cm]{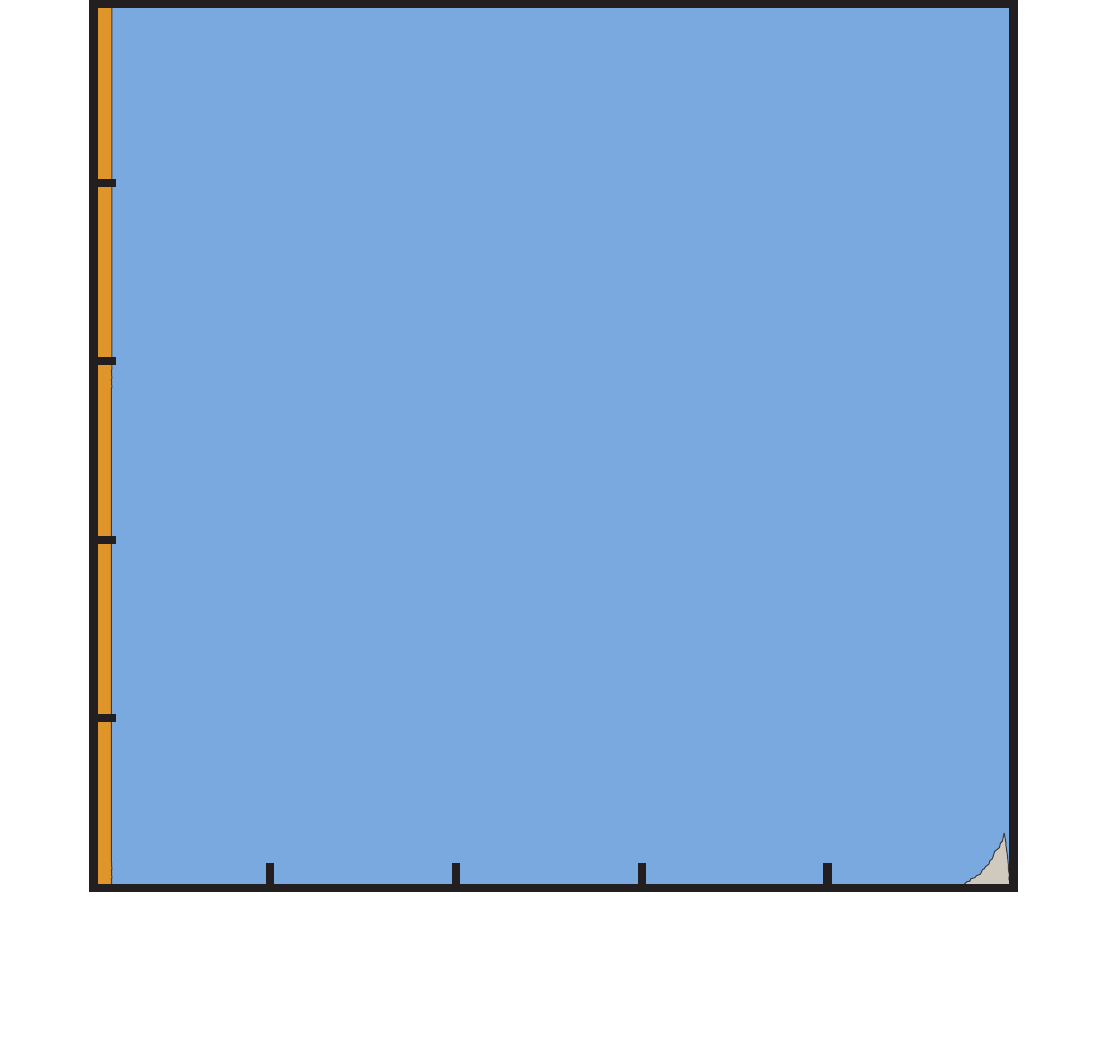}
\put(82,99){\scalebox{1.1}{(d)}}
\put(30,60){\scalebox{0.8}{\shortstack{\textbf{Mpemba}\\ ($|\mathcal{R}|<1,\ \kappa_0>1$)}}}
\put(-10,50){\rotatebox{90}{\scalebox{1.1}{$|\mathbf r_0|$}}}
\put(46,1){\scalebox{1.1}{$\theta/\pi$}}
\put(2,93){\scalebox{1.0}{1}}
\put(-2,77.5){\scalebox{1.0}{0.8}}
\put(-2,62){\scalebox{1.0}{0.6}}
\put(-2,45){\scalebox{1}{0.4}}
\put(-2,29){\scalebox{1}{0.2}}
\put(2,15){\scalebox{1}{0}}

\put(6.5,8){\scalebox{1}{0}}
\put(21,8){\scalebox{1}{$0.2$}}
\put(38.2,8){\scalebox{1}{$0.4$}}
\put(54.8,8){\scalebox{1}{$0.6$}}
\put(71.3,8){\scalebox{1}{$0.8$}}
\put(89.4,8){\scalebox{1}{$1$}}
\end{overpic}
\caption{
Robustness of the quantum Mpemba regime in a driven dissipative
two-level system.
(a),(b) Parameter dependence of the slow-mode amplitude ratio
$|\mathcal R|$ and the initial-distance ratio $\kappa_0$ in the
$(\Omega,\Delta)$ plane. White dashed curves mark the threshold
contours $|\mathcal R|=1$ and $\kappa_0=1$.
(c) Phase diagram obtained from the two criteria. For panels (a)--(c), the initial state is fixed at
\(|\mathbf r_0|=0.75\), \(\theta=\arccos(-0.8)\), and \(\phi=0\).
(d) Phase diagram in the $(\theta/\pi,|\mathbf r_0|)$ plane for
$\Omega=0.1$ and $\Delta=2.0$, with the azimuthal angle fixed at
$\phi=0$.
The blue region denotes the  quantum Mpemba regime,
$|\mathcal R|<1$ and $\kappa_0>1$; the red region denotes ordinary
acceleration, $|\mathcal R|<1$ and $\kappa_0\le1$; the light-gray region
has $\kappa_0>1$ but $|\mathcal R|\ge1$; and the yellow region has
$|\mathcal R|>1$ and $\kappa_0<1$.
Here $N=100$ and $T=0.1$; all other
parameters are the same as in Fig.~\ref{a2} of the main text.
}
 \label{Dis}
\end{figure}

{For a mixed input with \(|\mathbf r_0|=0.75\) and
\(\theta=\arccos(-0.8)\), slow-mode suppression persists over broad
parameter regions [Fig.~\ref{Dis}(a)], while the prepared state starts
farther from stationarity except in the strong-driving, small-detuning
region [Fig.~\ref{Dis}(b)]. Combining the two criteria gives the same
qualitative structure as in the pure-state case [Fig.~\ref{Dis}(c)].
In particular, a broad blue region with}
\(|\mathcal R|<1\) and \(\kappa_0>1\) persists, indicating that the
measurement-prepared state can start farther from the stationary state
with a reduced overlap with the slowest Liouvillian mode. 

{At fixed \(\Omega=0.1\) and \(\Delta=2.0\), the dependence on the
initial state is further tested by scanning its polar angle
\(\theta/\pi\) and Bloch length \(|\mathbf r_0|\)
[Fig.~\ref{Dis}(d)].} The blue region covers
most of the \((\theta/\pi,|\mathbf r_0|)\) plane, showing that the two
candidate criteria are not tied to a fine-tuned initial direction or to
a nearly pure state. The narrow yellow region corresponds to
\(|\mathcal R|>1\) and \(\kappa_0<1\), and therefore lies outside the
candidate region. These results demonstrate the persistence of the
slow-mode-suppression diagnostics against initial-state mixing.

\clearpage
\section{Fidelity and cumulative mixing in mixed-target steering}
\label{app:fidelity_coherence}

\begingroup
For a mixed target, the target population is replaced by the squared
Uhlmann fidelity. For qubit states with Bloch vectors $\mathbf u$ and
$\mathbf s$, this is
\begin{equation}
F(\mathbf u,\mathbf s)
=\frac12\left[1+\mathbf u\cdot\mathbf s
+\sqrt{(1-|\mathbf u|^2)(1-|\mathbf s|^2)}\right].
\label{eq:mixed_qubit_fidelity}
\end{equation}
We take $\rho_\tau=(I_2+s\sigma_z)/2$, with $0<s<1$.
Let $\theta_k$ denote successive Bloch-angle increments, beginning
at the input direction, and let $x$ be the final separation from
the effective target $\mathbf a_{\rm eff}=R_{\rm tot}^{-1}\hat{\mathbf z}$.
We vary $x$ within the geodesic family
\begin{equation}
\theta_1=\cdots=\theta_N
=\frac{\delta\theta_{\rm eff}-x}{N},
\qquad 0\le x\le\delta\theta_{\rm eff}.
\label{eq:mixed_target_family}
\end{equation}
Here $N\ge2$ and $0<\delta\theta_{\rm eff}<\pi$, so the first $N$
increments are smaller than $\pi/2$. Equal preceding increments are
an imposed control restriction, not a claim of global optimality for
mixed targets. For input Bloch length $r=|\mathbf r_0|$, the prepared
length, fidelity, and cumulative mixing are
\begin{align}
q_N(x)&=r\cos^N\!\left(\frac{\delta\theta_{\rm eff}-x}{N}\right),
\\
F_s(x)&=\frac12\left[1+s q_N(x)\cos x
+\sqrt{(1-s^2)[1-q_N(x)^2]}\right],
\label{eq:mixed_target_fidelity}
\\
\Sigma_{\rm mix}(x)&=
h_2\!\left(\frac{1+q_N(x)}{2}\right)
-h_2\!\left(\frac{1+r}{2}\right).
\label{eq:mixed_target_cost}
\end{align}
We minimize
\begin{equation}
\mathcal J_\lambda(x)=1-F_s(x)+\lambda\Sigma_{\rm mix}(x),
\qquad \lambda\ge0,
\label{eq:mixed_target_objective}
\end{equation}
within Eq.~(\ref{eq:mixed_target_family}). The penalty is the
cumulative coherence erased in the successive measurement bases,
as quantified by Eq.~(\ref{eq:Sigma_mix}), rather than the coherence
remaining in a fixed final basis.

\begin{figure}[!htb]
\centering
\includegraphics[width=0.98\textwidth]{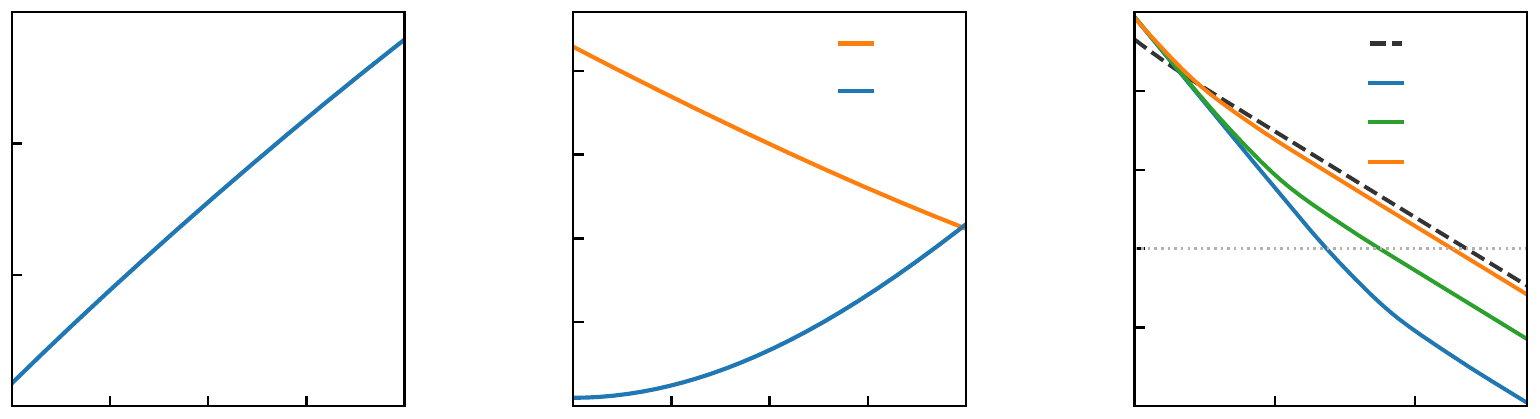}
\put(-383,136){\scalebox{1.1}{(a)}}
\put(-201,136){\scalebox{1.1}{(b)}}
\put(-20,136){\scalebox{1.1}{(c)}}
\put(-155,0){\scalebox{1}{$10^{-5}$}}
\put(-155,26){\scalebox{1}{$10^{-4}$}}
\put(-155,50){\scalebox{1}{$10^{-3}$}}
\put(-155,75){\scalebox{1}{$10^{-2}$}}
\put(-155,100){\scalebox{1}{$10^{-1}$}}
\put(-151,126){\scalebox{1}{$10^0$}}
\put(-132,-8){\scalebox{1}{0}}
\put(-88,-8){\scalebox{1}{$5$}}
\put(-44,-8){\scalebox{1}{$10$}}
\put(-10,-8){\scalebox{1}{$14$}}
\put(-350,40){\rotatebox{90}{\scalebox{1.0}{\text{Mixing/infidelity}}}}
\put(-60,-20){\scalebox{1.2}{$t$}}
\put(-42,120){\scalebox{0.9}{$\text{Unsteered}$}}
\put(-40,106){\scalebox{1}{$\lambda=0$}}
\put(-40,94){\scalebox{1}{$\lambda=0.5$}}
\put(-40,80){\scalebox{1}{$\lambda=2$}}
\put(-170,40){\rotatebox{90}{\scalebox{1.0}{$D_{\mathrm{T}}(\rho_t, \rho_{\mathrm{ss}})$}}}
\put(-326,0){\scalebox{1}{$0$}}
\put(-335,28){\scalebox{1}{$0.05$}}
\put(-332,54){\scalebox{1}{$0.1$}}
\put(-335,81){\scalebox{1}{$0.15$}}
\put(-332,110){\scalebox{1}{$0.2$}}
\put(-335,127){\scalebox{1}{$0.24$}}
\put(-315,-8){\scalebox{1}{0}}
\put(-288,-8){\scalebox{1}{$0.5$}}
\put(-252,-8){\scalebox{1}{$1$}}
\put(-225,-8){\scalebox{1}{$1.5$}}
\put(-190,-8){\scalebox{1}{$2$}}
\put(-350,40){\rotatebox{90}{\scalebox{1.0}{\text{Mixing/infidelity}}}}
\put(-252,-20){\scalebox{1.2}{$\lambda$}}
\put(-213,118){\scalebox{1}{$\Sigma_{\rm mix}$}}
\put(-213,104){\scalebox{1}{$1-F_s$}}
\put(-506,0){\scalebox{1}{$0$}}
\put(-512,43){\scalebox{1}{$0.3$}}
\put(-512,86){\scalebox{1}{$0.6$}}
\put(-512,127){\scalebox{1}{$0.9$}}
\put(-498,-8){\scalebox{1}{0}}
\put(-471,-8){\scalebox{1}{$0.5$}}
\put(-435,-8){\scalebox{1}{$1$}}
\put(-408,-8){\scalebox{1}{$1.5$}}
\put(-372,-8){\scalebox{1}{$2$}}
\put(-520,55){\rotatebox{90}{\scalebox{1.0}{$x^*$(rad)}}}
\put(-435,-20){\scalebox{1.2}{$\lambda$}}
\vspace{6pt}
\caption{{{Mixed-target steering with a cumulative mixing penalty.
(a) Optimal terminal angle $x^*$.
(b) Mixing $\Sigma_{\rm mix}$ and infidelity $1-F_s$.
(c) Subsequent trace-distance relaxation; the dotted line marks
$D_T=10^{-3}$. Parameters: $r=0.9$, $s=0.8$, $N=10$,
$\theta_0=\arccos(-0.6)$, $\phi_0=0$, $\Omega=0.1$, $\Delta=2$,
$T=0.1$, and $\gamma=1$. Time excludes preparation.}}}
\label{fig:mixed_target_cost}
\end{figure}

Figure~\ref{fig:mixed_target_cost} uses a mixed input $r=0.9$ and
a mixed target $s=0.8$. Bounded scalar minimization was checked on
a grid of $100001$ angles for each of $201$ equally spaced weights
in $0\le\lambda\le2$. Over this range, increasing $\lambda$ shifts the optimal terminal angle
farther from the target [Fig.~\ref{fig:mixed_target_cost}(a)] and
reduces mixing at the expense of fidelity
[Fig.~\ref{fig:mixed_target_cost}(b)]. At $\lambda=0,0.5,2$, respectively,
the optimal fidelities are $0.9950,0.9876,0.8916$, and the mixing
costs are $0.2142,0.1845,0.1058$ (natural logarithms). All input,
target, and dynamical parameters remain fixed.

Fidelity--mixing optimization alone does not establish a Mpemba
effect. We therefore propagate the three optimized states under the
same dissipative qubit dynamics [Fig.~\ref{fig:mixed_target_cost}(c)].
Their initial-distance ratios are $\kappa_0=1.9145,1.9388,1.8780$,
and their relaxation times to $D_T=10^{-3}$ are
$ t=6.8422,8.7304,11.3130$, compared with $11.8262$
for the unsteered input. All three distance curves cross the
unsteered curve and reach this tolerance earlier despite starting
farther from stationarity. These verified examples exclude
preparation time and apparatus costs.

For a general finite-dimensional target, the fidelity is
\begin{equation}
F(\tilde\rho_N,\rho_\tau)
=\left[\mathrm{Tr}\sqrt{
\sqrt{\rho_\tau}\tilde\rho_N\sqrt{\rho_\tau}}\right]^2
=\left(\sum_j\sqrt{\nu_j}\right)^2,
\end{equation}
where $\nu_j$ are the eigenvalues of
$\sqrt{\rho_\tau}\tilde\rho_N\sqrt{\rho_\tau}$.
Implementation-dependent resources can be included through
\begin{equation}
\mathcal J=1-F(\tilde\rho_N,\rho_\tau)
+\lambda\Sigma_{\rm mix}+\eta\mathcal C_{\rm ctrl},
\qquad\lambda,\eta\ge0.
\end{equation}
The control cost $\mathcal C_{\rm ctrl}$ and its units require a
specified physical implementation; they cannot be inferred from
$\Sigma_{\rm mix}$. This defines a higher-dimensional optimization
problem, without asserting a general optimal schedule or a
relaxation advantage that has not been checked independently.

\endgroup

\twocolumngrid

\end{document}